\documentclass[prd,aps,showpacs,notitlepage,superscriptaddress,nofootinbib]{revtex4-1}

\usepackage{graphicx}
\usepackage{color}
\usepackage{url}
\usepackage{amsmath}
\usepackage{multirow}

\def\bea#1\eea{\begin{align}#1\end{align}}

\makeatletter
\def\slash#1{{\mathpalette\c@ncel{#1}}} 
\makeatother

\newcommand\beq{\begin{eqnarray}}
\newcommand\eeq{\end{eqnarray}}

\newcommand\la{\langle}
\newcommand\ra{\rangle}

\begin{document}
\title{Reformulation of the twist-3 gluon Sivers effect toward the application to 
the heavy quarkonium production}

\date{\today}

\author{Shinsuke Yoshida}
\email{shinyoshida85@gmail.com}
\affiliation{Guangdong Provincial Key Laboratory of Nuclear Science, Institute of Quantum Matter, South
China Normal University, Guangzhou 510006, China}
\affiliation{Guangdong-Hong Kong Joint Laboratory of Quantum Matter, Southern Nuclear Science Computing Center, South China Normal University, Guangzhou 510006, China}

\author{Difei Zheng}
\email{difeizheng@163.com}
\affiliation{Guangdong Provincial Key Laboratory of Nuclear Science, Institute of Quantum Matter, South
China Normal University, Guangzhou 510006, China} 
\affiliation{Guangdong-Hong Kong Joint Laboratory of Quantum Matter, Southern Nuclear Science Computing Center, South China Normal University, Guangzhou 510006, China}
\affiliation{School of  Physics and Telecommunication Engineering, Guangdong Provincial Engineering Research Center for Optoelectronic Instrument, South China Normal University, Guangzhou 510006, China}

\begin{abstract}

In this paper, we propose a new calculation method for the twist-3 gluon Sivers effect 
within the collinear factorization approach. The method called pole calculation has been used to 
derive the cross section formula for the single transverse-spin asymmetry(SSA) as 
a standard method. We
point out that we encounter a problem when we try to apply the pole calculation to the SSA
in the heavy quarkonium production
whose hadronization mechanism is described by non-relativistic QCD(NRQCD) framework. 
We show that the new 
calculation method solves this problem and successfully reproduces known results derived 
by the pole calculation. Our new method extends the applicability of the twist-3 calculation technique
to heavy quarkonium productions which are ideal observables for the investigation of 
the gluon Sivers effect.

\end{abstract}

\maketitle

\section{Introduction}

Unveiling a novel structure of hadron through the study of the single transverse-spin asymmetry(SSA)
has been one of major research directions in high energy hadron physics in recent decades. 
Much experimental effort has been devoted to the measurement of SSAs in many different processes 
and different kinematic regions since the first observation of large SSAs in the late 70s\cite{Klem:1976ui,Bunce:1976yb}.
Sivers effect generated by a transversely polarized proton 
is a possible source of the large SSA and it has drawn much attention in the context of the 
nucleon structure because it gives a new perspective on the orbital spin structure inside the proton. 
The quark Sivers effect has been well understood in the past couple of decades through the study of 
SSAs in many processes\cite{Anselmino:2008sga,Anselmino:2012aa,Anselmino:2013rya,Echevarria:2014xaa,Martin:2017yms,Echevarria:2020hpy,Bury:2020vhj,Bury:2021sue}. On the other hand, the gluon Sivers effect has been hardly 
understood because available data is limited. 
The understanding of the gluon Sivers effect is the central problem 
in the high energy spin physics.

Electron-ion collider(EIC), the next-generation collider
experiment, could measure SSAs in many processes including heavy flavored meson productions 
which are ideal observables for the investigation of  
the gluon Sivers effect. The project of EIC has motivated 
much theoretical work on the development of perturbative QCD techniques in recent years.  
Two perturbative QCD frameworks have been mainly developed 
in order to study the origin of the large SSA in different kinematic regions. 
The transverse-momentum-dependent(TMD) factorization is valid in the 
low transverse momentum region of a produced hadron, and
in contrast the collinear twist-3 factorization is applicable in the high transverse momentum region.
The EIC experiment is expected to measure SSAs in a wide range of the transverse momentum 
as the RHIC experiment has done. Developments of both two frameworks are required for the complete understanding of the SSAs measured in EIC.
The SSA in the heavy charmonium production is one of desired observables for the investigation of 
the gluon Sivers effect because a heavy quark pair is mainly produced by the gluon fusion process.
The gluon Sivers effect on the $J/\psi$ SSA has been well discussed within the TMD factorization
\cite{Godbole:2014tha,Mukherjee:2016qxa,DAlesio:2017rzj,Rajesh:2018qks,DAlesio:2018rnv,Sun:2019tuk,DAlesio:2019qpk,Kishore:2019fzb,DAlesio:2019gnu,DAlesio:2020eqo}.
However, no calculation has been done within the collinear twist-3 factorization although the basic technique was already formulated in $D$-meson production \cite{Kang:2008qh,Beppu:2010qn}. 
The main reason why it has not been 
done yet is that there is a conflict
between the conventional twist-3 calculation for the gluon Sivers effect 
and non-relativistic QCD(NRQCD) framework which describes the hadronization mechanism of
$J/\psi$. The purpose of this paper is that we resolve the conflict by proposing a new calculation
method for the twist-3 gluon Sivers effect. We will show that the new method can reproduce 
known results derived by the conventional calculation in \cite{Beppu:2010qn} and does not have any conflicts with the NRQCD framework.
Our result in this paper will give a theoretical basis to the application of the twist-3 framework 
to SSAs in heavy quarkonium productions.

The remainder of this paper is organized as follows:
In section II, we introduce definitions of the twist-3 gluon distribution functions relevant to 
the present study and show some relations  
among them.  
In section III, we briefly review the conventional pole calculation and show the problem caused 
when we apply the pole calculation to the $J/\psi$ SSA.
In section IV (and Appendix A), we show the new calculation technique for twist-3 gluon Sivers effect in detail.
Section V is devoted to a summary of the present study.


\section{Definitions of the twist-3 gluon functions for the transversely polarized
proton}

The twist-3 cross section is in general expressed by three types of nonperturbative functions\cite{Kanazawa:2015ajw}. 
Two of them, kinematical and dynamical
functions, are relevant in the case of the proton Sivers effect. Here we recall definitions of those two types of functions for the gluon and some relations between them derived in \cite{Koike:2019zxc}. 

\

\noindent
(1) kinematical twist-3 gluon distribution functions

\

Kinematicial functions are often referred to as the first $k^2_T/M^2$-moment of corresponding TMD functions\cite{Mulders:2000sh}.
Exact definitions of the kinematical twist-3 gluon distribution functions are given by
\beq
\Phi^{\alpha\beta\gamma}_{\partial}(x)&=&\int{d\lambda\over 2\pi}e^{i\lambda x}
\la pS_{\perp}|F^{\beta n}
(0)F^{\alpha n}(\lambda n)|pS_{\perp}\ra(i\overleftarrow{\partial}^{\gamma}_{\perp})
\nonumber\\
&\equiv&\lim_{\xi_{\perp}\to 0}\int{d\lambda\over 2\pi}e^{i\lambda x}\la pS_{\perp}|
\Bigl(F^{\beta n}(0)[0,\infty n]\Bigr)_ai{d\over d\xi_{\perp\gamma}}
\Bigl([\infty n+\xi_{\perp},\lambda n+\xi_{\perp}]F^{\alpha n}(\lambda n+\xi_{\perp})\Bigr)_a|pS_{\perp}\ra
\nonumber\\
&=&{M_N\over 2}g^{\alpha\beta}_{\perp}\epsilon^{pnS_{\perp}\gamma}G^{(1)}_T(x)
+i{M_N\over 2}\epsilon^{pn \alpha\beta}S_{\perp}^{\gamma}\Delta G^{(1)}_T(x)
+{M_N\over 8}\Bigl(\epsilon^{pnS_{\perp}\{\alpha}g^{\beta\}\gamma}_{\perp}
+\epsilon^{pn\gamma\{\alpha}S^{\beta\}}_{\perp}\Bigr)\Delta H^{(1)}_{T}(x)+\cdots,
\label{kinematical}
\eeq
where $p, S_{\perp}$ and $M_N$ represent the proton's momentum, spin and mass respectively. 
We use a shorthand notation 
$\epsilon^{pnS_{\perp}\gamma}=\epsilon^{\mu\nu\rho\gamma}p_{\mu}n_{\nu}S_{\perp\rho}$ throughout 
this paper. 
$[0,\lambda n]$ denotes the gauge-link operator 
$[0,\lambda n]\equiv P\exp\Bigl(ig\int_{\lambda}^0dt\, A^n(tn)\Big)$ which guarantees 
the gauge-invariance of the matrix element.
$n$ is a light-like vector satisfying $p\cdot n=1, n^2=0$. The twist-3 functions $G^{(1)}_T(x)$ 
and $\Delta H^{(1)}_{T}(x)$ are relevant to the study of the SSA, whereas the contribution from 
$\Delta G^{(1)}_T(x)$ is canceled because it gives a pure imaginary contribution.

\

\noindent
(2) Dynamical twist-3 gluon distribution functions

\

Dynamical twist-3 gluon distribution functions are defined by a matrix element of three gluon field 
strength tensors.  There are two different types, $C$-even function $N(x_1,x_2)$ and 
$C$-odd function $O(x_1,x_2)$, reflecting the fact that there are two structure constants $if^{abc}$
and $d^{abc}$ in SU($N_c$) group,
\beq
N^{\alpha\beta\gamma}(x_1,x_2)&=&i\int{d\lambda\over 2\pi}\int{d\mu\over 2\pi}
e^{i\lambda x_1}e^{i\mu (x_2-x_1)}
\la pS_{\perp}|if^{bca}F_{b}^{\beta n}(0)
F_c^{\gamma n}(\mu n)F_a^{\alpha n}(\lambda n)|pS_{\perp}\ra
\nonumber\\
&=&2iM_N\Bigl[g_{\perp}^{\alpha\beta}\epsilon^{\gamma pnS_{\perp}}N(x_1,x_2)
-g_{\perp}^{\beta\gamma}\epsilon^{\alpha pnS_{\perp}}N(x_2,x_2-x_1)
-g_{\perp}^{\alpha\gamma}\epsilon^{\beta pnS_{\perp}}N(x_1,x_1-x_2)\Bigr]+\cdots,
\label{C-even}
\eeq
\beq
O^{\alpha\beta\gamma}(x_1,x_2)&=&i\int{d\lambda\over 2\pi}\int{d\mu\over 2\pi}
e^{i\lambda x_1}e^{i\mu (x_2-x_1)}
\la pS_{\perp}|d^{bca}F_{b}^{\beta n}(0)
F_c^{\gamma n}(\mu n)F_a^{\alpha n}(\lambda n)|pS_{\perp}\ra
\nonumber\\
&=&2iM_N\Bigl[g_{\perp}^{\alpha\beta}\epsilon^{\gamma pnS_{\perp}}O(x_1,x_2)
+g_{\perp}^{\beta\gamma}\epsilon^{\alpha pnS_{\perp}}O(x_2,x_2-x_1)
+g_{\perp}^{\alpha\gamma}\epsilon^{\beta pnS_{\perp}}O(x_1,x_1-x_2)\Bigr]+\cdots,
\label{C-odd}
\eeq
where we omitted gauge-links for simplicity. These dynamical functions have the following symmetries.
\beq
O(x_1,x_2)&=&O(x_2,x_1),\hspace{5mm}O(x_1,x_2)=O(-x_1,-x_2),
\nonumber\\
N(x_1,x_2)&=&N(x_2,x_1),\hspace{5mm}N(x_1,x_2)=-N(-x_1,-x_2).
\label{symmetries}
\eeq
Each kinematical function has a relation with the $C$-even dynamical function 
by the equivalence between matrix elements (\ref{kinematical}) and (\ref{C-even}) as
\beq
G_T^{(1)}(x)=-4\pi(N(x,x)-N(x,0)),\hspace{5mm}\Delta H_T^{(1)}(x)=8\pi N(x,0).
\label{relations}
\eeq
We will show that these relations are required to reproduce known results calculated by the conventional pole calculation.


\section{Pole calculation}


\subsection{Introduction to the conventional pole calculation}

We introduce the conventional pole calculation in order to compare it with the
new method in the next section and to see a problem in applying it to $J/\psi$ production.
It is known that a na\"{\i}vely $T$-odd observable like the SSA has to be canceled without a nontrivial
imaginary phase. When we apply the collinear framework to the proton Sivers effect, the imaginary phase is given by the imaginary part of a propagator. A parton propagator with the momentum $k$ is given by
\beq
{1\over k^2+i\epsilon}=P{1\over k^2}-i\pi\delta(k^2).
\label{pole_separation}
\eeq
The second term with the delta function gives the imaginary phase and this is called 
pole contribution. In the conventional pole calculation, we replace a propagator with the delta function
term at the beginning of the calculation.
We here briefly review the pole calculation in semi-inclusive deep 
inelastic scattering(SIDIS),
\beq
e(\ell)+p^{\uparrow}(p,S_{\perp})\to e(\ell')+\pi(P_h)+X.
\eeq
The polarized cross section formula for the pion production is given 
in the hadron frame \cite{Koike:2011ns} as
\beq
\frac{d^{6}\Delta\sigma}{dx_{bj}dQ^{2}dz_{f}dP^{2}_{h}d\phi d\chi}  =
\frac{\alpha ^{2}_{em}}{128\pi^4 z_fS^{2}_{ep} x^{2}_{bj} Q^{2}}
L^{\rho\sigma}(\ell,\ell')W_{\rho\sigma}(p,q,P_h), 
\eeq
where $\alpha_{em}=e^2/4\pi$ is the QED coupling constant, 
$\chi$ and $\phi$ are the azimuthal angles of the hadron plane and 
the lepton plane respectively, the leptonic tensor is given by 
$L^{\rho\sigma}(\ell,\ell')=2(\ell^\rho \ell'^\sigma+\ell^\sigma \ell'^{\rho})
-Q^2g^{\rho\sigma}$ and we use the following Lorentz invariant variables.
\beq
S_{ep}&=&(p+\ell)^2,\hspace{5mm}
Q^2=-q^2=-(\ell-\ell')^2,\hspace{5mm}
x_{bj}=\frac{Q^2}{2p\cdot q},\hspace{5mm}
z_f=\frac{p\cdot P_h}{p\cdot q}.
\eeq
The hadronic tensor $W_{\rho\sigma}(p,q,P_h)$ describes the scattering of the proton and 
the virtual photon. The twist-3 effect from the proton is convoluted with the parton fragmentation
function of the pion $D(z)$,
\beq
W_{\rho\sigma}(p,q,P_h)=\int{dz\over z^2}D(z)w_{\rho\sigma}(p,q,{P_h\over z}).
\eeq
We show how to calculate the twist-3 contribution from
$w_{\rho\sigma}(p,q,{P_h\over z})$ in the conventional method.
The imaginary phase in (\ref{pole_separation}) requires 
an interference with one coherent gluon as shown in FIG. 1.
\begin{figure}[h]
\begin{center}
  \includegraphics[height=6cm,width=12cm]{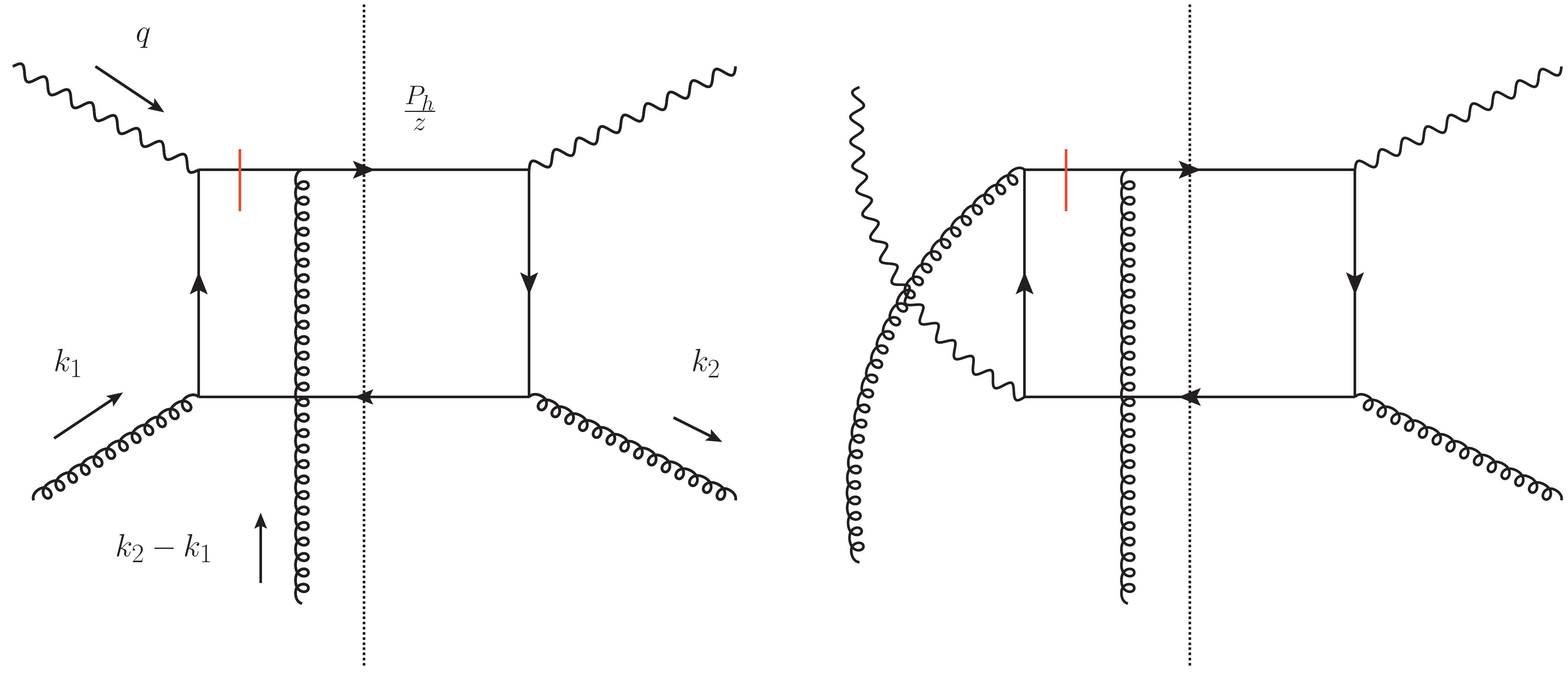}\vspace{2mm}

  \includegraphics[height=6cm,width=12cm]{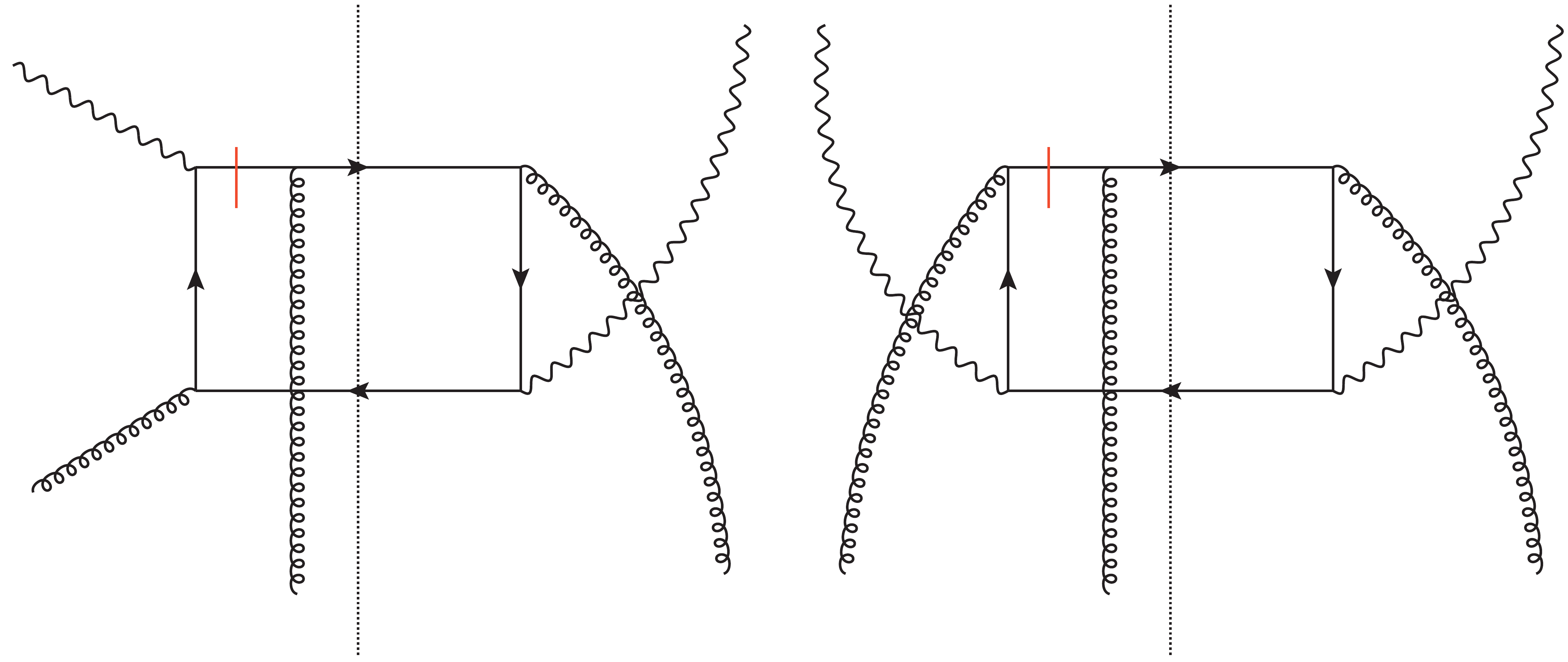}
\end{center}
 \caption{$S^{abc}_{\mu\nu\lambda}(k_1,k_2)$ is given by the sum of these diagrams and
their complex conjugate diagrams. The coherent gluon with the momentum $k_2-k_1$ generates
the additional propagator indicated by the red bar. The red barred propagator is replaced with the
delta function $-i\pi\delta\Bigl([{P_h\over z}-(k_2-k_1)]^2\Bigr)$ by using (\ref{pole_separation}).}
\label{fig1}
\end{figure}
The coherent gluon line generates the additional propagator indicated by the red bar and this 
propagator is replaced with the delta function by using the formula (\ref{pole_separation}). The corresponding contribution to FIG. 1 is mathematically given by
\beq
w_{\rho\sigma}(p,q,{P_h\over z})=\int{d^4k_1\over (2\pi)^4}\int{d^4k_2\over (2\pi)^4}
\int{d^4\xi}\int{d^4\eta}\,e^{ik_1\cdot \xi}e^{i\eta\cdot(k_2-k_1)}
\la pS_{\perp}|A_b^{\nu}(0)gA_c^{\lambda}(\eta)A_a^{\mu}(\xi)|pS_{\perp}\ra
S^{abc}_{\mu\nu\lambda, \rho\sigma}(k_1,k_2).
\label{pole_calculation}
\eeq
The hard part $S^{abc}_{\mu\nu\lambda, \rho\sigma}(k_1,k_2)$ is given by the sum of the diagrams in FIG. 1.
We omit the virtual photon indices $\rho,\sigma$ for simplicity.
The proton matrix element consists of three gluon fields
which are not gauge invariant and $w(p,q,{P_h\over z})$
includes all twist effects. 
A systematic way to extract the twist-3 contribution from (\ref{pole_calculation}) was developed 
in \cite{Beppu:2010qn} and we can derive the gauge-invariant expression as
\beq
&&w(p,q,{P_h\over z})
\nonumber\\
&=&\omega^{\mu}_{\ \alpha}\omega^{\nu}_{\ \beta}
\omega^{\lambda}_{\ \gamma}\int{dx_1\over x_1}\int{dx_2\over x_2}
\Bigl({-if^{abc}\over N_c(N_c^2-1)}N^{\alpha\beta\gamma}(x_1,x_2)
+{N_cd^{abc}\over (N_c^2-4)(N_c^2-1)}O^{\alpha\beta\gamma}(x_1,x_2)\Bigr){\partial\over \partial k_2^{\lambda}}
S^{abc}_{\mu\nu p}(k_1,k_2)\Bigr|_{k_i=x_ip},\hspace{7mm} 
\label{pole_result}
\eeq
where $\omega^{\mu}_{\ \nu}=g^{\mu}_{\ \nu}-p^{\mu}n_{\nu}$.
A reader can use this formula without following the derivation in detail if the hard part
$S^{abc}_{\mu\nu\lambda}(k_1,k_2)$ satisfies four required conditions. 
Three of them are Ward-Takahashi identities(WTIs),
\beq
(k_2-k_1)^{\lambda}S^{abc}_{\mu\nu\lambda}(k_1,k_2)&=&0,
\label{WTI1}
\\
k_1^{\mu}S^{abc}_{\mu\nu\lambda}(k_1,k_2)&=&0,
\\
k_2^{\nu}S^{abc}_{\mu\nu\lambda}(k_1,k_2)&=&0.
\label{WTI3}
\eeq
One can easily check that the diagrams in FIG. 1 satisfy these conditions by 
taking into account the delta function
given by the pole contribution. There is one more required condition 
\beq
\Bigl({\partial\over \partial k_2^{\lambda}}S^{abc}_{\mu\nu p}(k_1,k_2)+
{\partial\over \partial k_1^{\lambda}}S^{abc}_{\mu\nu p}(k_1,k_2)\Bigr)\Bigr|_{k_i=x_ip}=0.
\label{additional}
\eeq
Note that this condition cannot be derived by taking the derivative of
$(k_2-k_1)^{\lambda}S^{abc}_{\mu\nu\lambda}(k_1,k_2)=0$ with respect to
$k_{1,2}$
because the pole contribution gives $x_1=x_2$ in the collinear limit. 
Although the WTIs (\ref{WTI1}) - (\ref{WTI3}) are followed by the gauge-invariance,
the condition (\ref{additional}) is not clearly based on such a symmetry of the field theory.
Therefore, we have to verify that the sum of diagrams satisfies the condition (\ref{additional}) case by case
with a direct inspection when we apply the pole method. 
\begin{figure}[h]
\begin{center}
  \includegraphics[height=6cm,width=12cm]{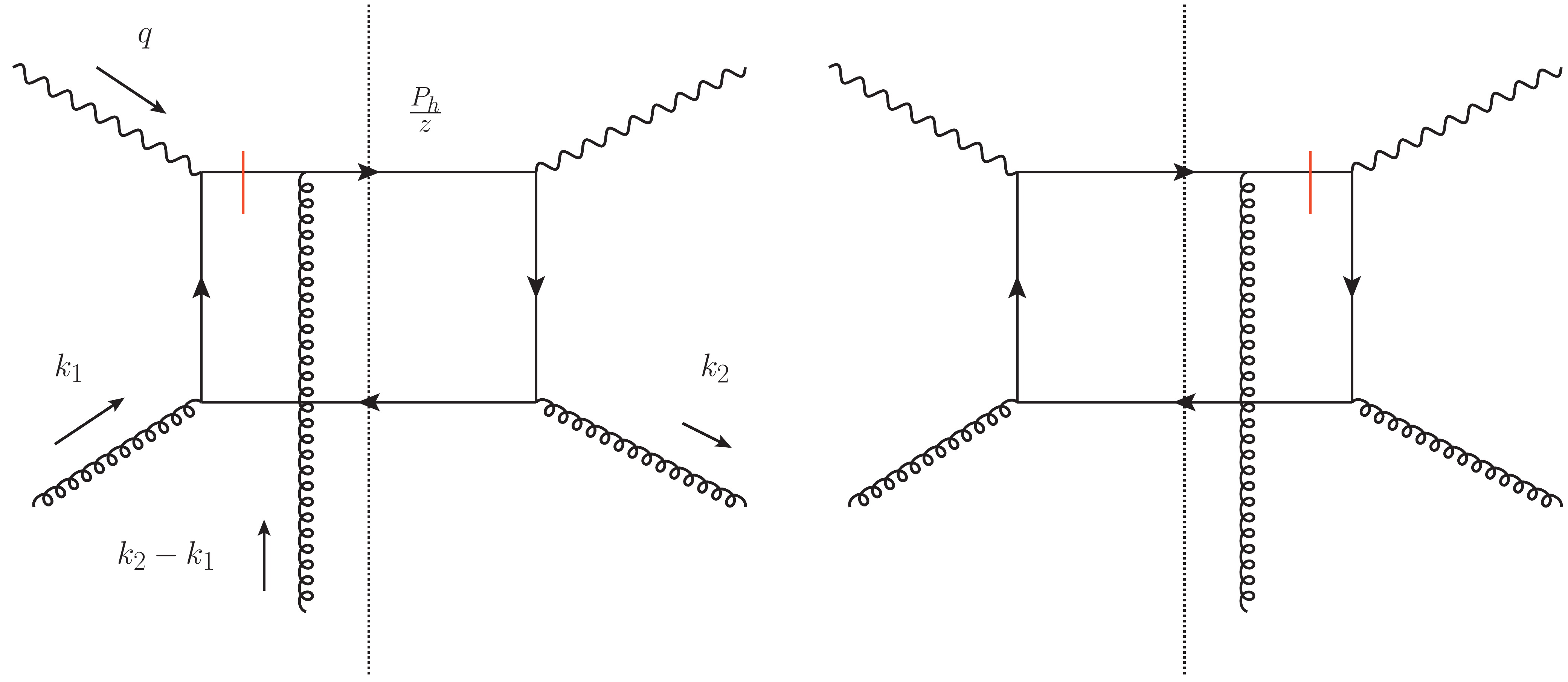}
\end{center}
 \caption{A pair of the diagram which satisfies the condition (\ref{additional}).}
\label{fig1}
\end{figure}
Here we check that the pair of diagrams in FIG. 2 satisfies the condition by introducing 
the explicit form as follows:
\beq
S^{abc}_{\mu\nu p}(k_1,k_2)\Bigr|_{\rm FIG. 2}&=&
-{\rm Tr}[{\slash{P}_h\over z}\slash{p}\Bigl({\slash{P}_h\over z}-(\slash{k}_2-\slash{k}_1)\Bigr)
\gamma_{\rho}\Bigl({\slash{P}_h\over z}-(\slash{k}_2-\slash{k}_1)-\slash{q}\Bigr)\gamma_{\mu}
(\slash{k}_2+\slash{q}-{\slash{P}_h\over z})\gamma_{\nu}
({\slash{P}_h\over z}-\slash{q})\gamma_{\rho}]
\nonumber\\
&&\times {1\over [{P_h\over z}-(k_2-k_1)-q]^2}{1\over ({P_h\over z}-q)^2}
\Bigl[-i\pi\delta\Bigl([{P_h\over z}-(k_2-k_1)]^2\Bigr)\Bigr]
2\pi\delta\Bigl((k_2+q-{P_h\over z})^2\Bigr)
\nonumber\\
&&-{\rm Tr}[{\slash{P}_h\over z}
\gamma_{\rho}\Bigl({\slash{P}_h\over z}-\slash{q}\Bigr)\gamma_{\mu}
(\slash{k}_1+\slash{q}-{\slash{P}_h\over z})\gamma_{\nu}
\Bigl({\slash{P}_h\over z}+(\slash{k}_2-\slash{k}_1)-\slash{q}\Bigr)\gamma_{\rho}
\Bigl({\slash{P}_h\over z}+(\slash{k}_2-\slash{k}_1)\Bigr)\slash{p}]
\nonumber\\
&&\times {1\over [{P_h\over z}+(k_2-k_1)-q]^2}{1\over ({P_h\over z}-q)^2}
\Bigl[i\pi\delta\Bigl([{P_h\over z}+(k_2-k_1)]^2\Bigr)\Bigr]
2\pi\delta\Bigl((k_1+q-{P_h\over z})^2\Bigr),
\label{example}
\eeq
where we omitted the common color factor. We can find that two diagrams are canceled at $x_1=x_2$
in the collinear limit
because they are topologically the same and the imaginary phases
caused by (\ref{pole_separation}) have opposite signs.
The $(k_2-k_1)$-dependent parts obviously satisfies (\ref{additional}). 
We find that the terms given by $k_2$-derivative acting on
$(k_2+q-{P_h\over z})$-dependence are canceled with
the terms given by $k_1$-derivative acting on $(k_1+q-{P_h\over z})$-dependence
in the collinear limit $x_1=x_2$, which means 
the condition (\ref{additional}) is satisfied.
One can check that other pairs in FIG. 1 satisfy (\ref{additional}) in the same way. 
We have confirmed that (\ref{additional}) is still satisfied when we take into account a heavy quark mass
\cite{Beppu:2010qn} as long as an amplitude and its complex conjugate are connected by a single trace factor 
Tr[$\cdots$] like (\ref{example}). 
As stated above, the condition (\ref{additional}) is not based on fundamental symmetries of the theory and
it is not guaranteed to hold in the general case.
As we see in the next subsection, NRQCD is a counter-example to (\ref{additional})
and it hinders an application of the pole calculation to the SSA in the heavy quarkonium production
when we use NRQCD framework to describe the
hadronization mechanism. Elimination of the condition (\ref{additional}) extends 
the applicability of the twist-3 calculation method which is necessary for the understanding of SSAs in
heavy quarkonium productions.


\subsection{A problem in the extension to the heavy quarkonium production}

NRQCD is a widely accepted theoretical framework for the description of the hadronization mechanism of 
a heavy quarkonium\cite{Caswell:1985ui,Bodwin:1994jh}. The $J/\psi$ production in SIDIS is illustrated as
\beq
e(\ell)+p^{\uparrow}(p,S_{\perp})\to e(\ell')+\sum_nc\bar{c}[n](P_h)+X,
\eeq 
where $n={}^3S_1^{[1]},{}^1S_0^{[8]},{}^3S_1^{[8]},\cdots$ 
denotes possible Fock states of the charm quark pair hadronizing into $J/\psi$. Here we focus on the
color-singlet contribution(${}^3S_1^{[1]}$) to the $J/\psi$ production as one of the examples so that one can see the breakdown of the condition (\ref{additional}). A typical diagram which gives the color-singlet
contribution is shown in FIG. 3. 
\begin{figure}[h]
\begin{center}
  \includegraphics[height=6cm,width=10cm]{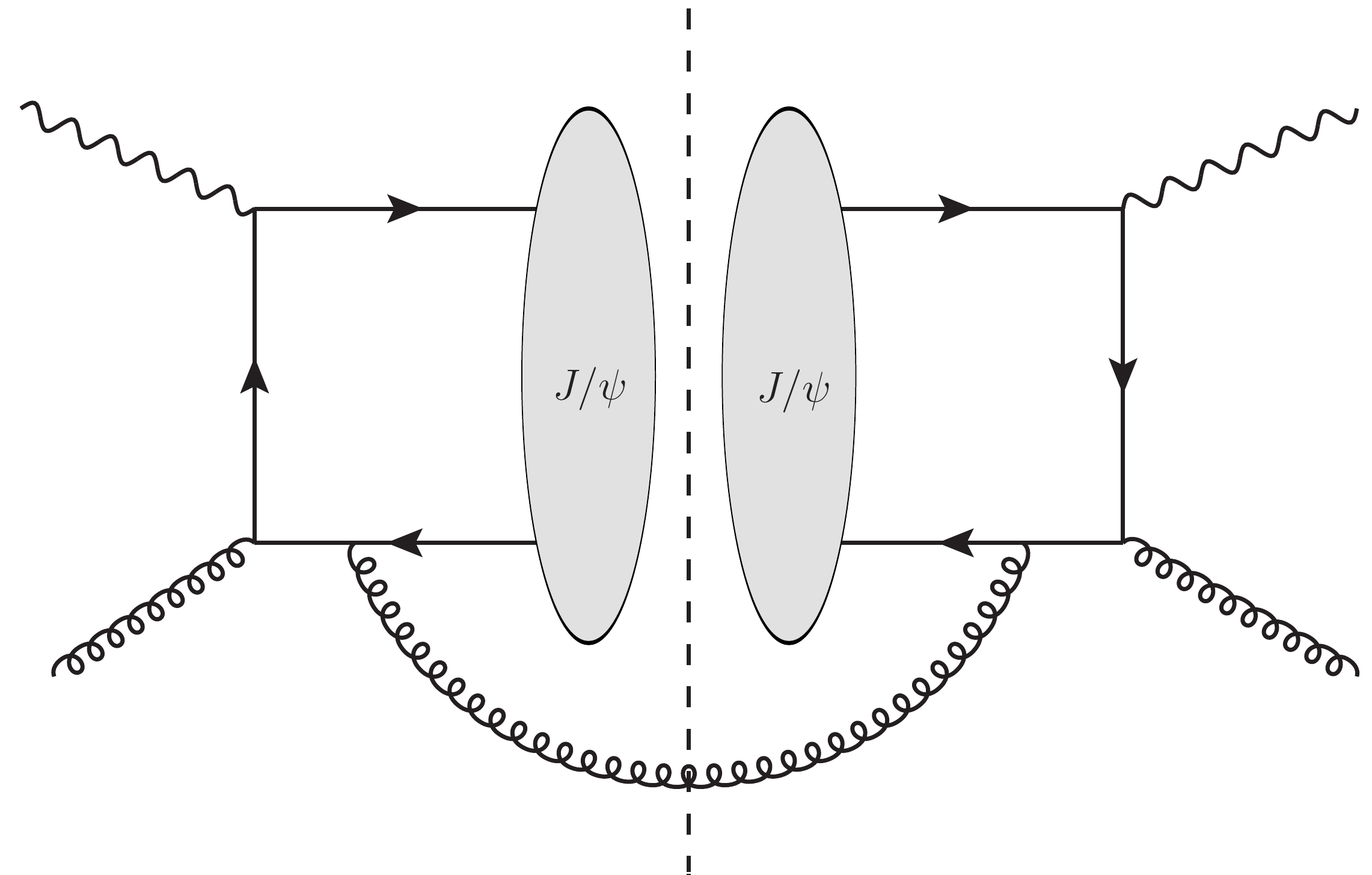}
\end{center}
 \caption{A typical diagram which gives the color-singlet contribution.}
\end{figure}
We insert the coherent gluon line to the diagram in order to give the imaginary phase
as shown in FIG. 4.
\begin{figure}[h]
\begin{center}
  \includegraphics[height=6cm,width=16cm]{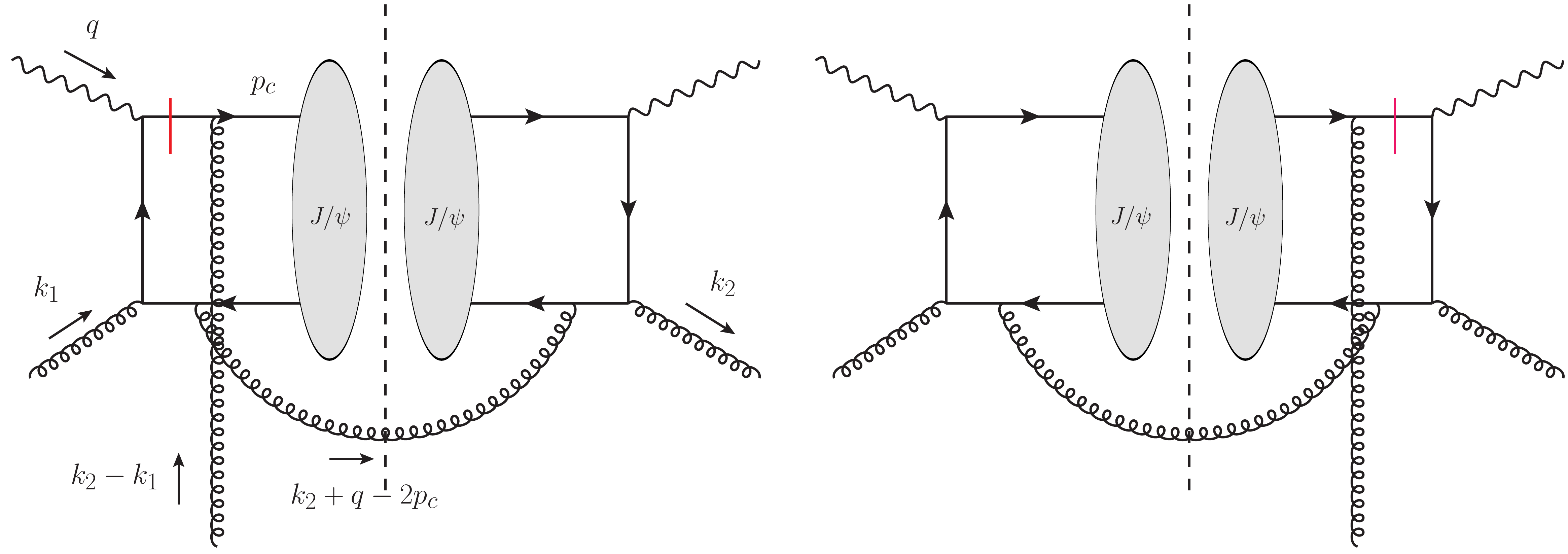}
\end{center}
 \caption{Corresponding pair of diagrams to the pair in FIG. 2.}
\end{figure}
The condition (\ref{additional}) 
should be satisfied in the pair of diagrams in FIG. 4 in analogy with the pion production.
We can show the explicit form as follows:
\beq
S^{abc}_{\mu\nu p}(k_1,k_2)\Bigr|_{\rm FIG. 4}&=&
-{\rm Tr}[\Pi_1^{\alpha}\slash{p}\Bigl(\slash{p}_c-(\slash{k}_2-\slash{k}_1)+m_c\Bigr)
\gamma_{\rho}\Bigl(\slash{p}_c-(\slash{k}_2-\slash{k}_1)-\slash{q}+m_c\Bigr)
\gamma_{\mu}(\slash{p}_c-\slash{k}_2-\slash{q}+m_c)\gamma_{\tau}]
\nonumber\\
&&\times{\rm Tr}[\Pi_1^{\dagger\beta}\gamma_{\delta}(\slash{p}_c-\slash{k}_2-\slash{q}+m_c)
\gamma_{\nu}(\slash{p}_c-\slash{q}+m_c)
\gamma_{\sigma}]\Bigl[-i\pi\delta\Bigl([p_c-(k_2-k_1)]^2-m_c^2\Bigr)\Bigr]
\nonumber\\
&&\times{1\over [p_c-(k_2-k_1)-q]^2-m_c^2}
{1\over [(p_c-k_2-q)^2-m_c^2]^2}{1\over (p_c-q)^2-m_c^2}\Pi_{\alpha\beta}
g^{\tau\delta}_{\perp}(k_2+q-2p_c)
\nonumber\\
&&\times\delta\Bigl((k_2+q-2p_c)^2\Bigr)
\nonumber\\
&&-{\rm Tr}[\Pi_1^{\alpha}\gamma_{\rho}\Bigl(\slash{p}_c-\slash{q}+m_c\Bigr)
\gamma_{\mu}(\slash{p}_c-\slash{k}_1-\slash{q}+m_c)\gamma_{\tau}]
\Bigl[i\pi\delta\Bigl([p_c+(k_2-k_1)]^2-m_c^2\Bigr)\Bigr]
\nonumber\\
&&\times{\rm Tr}[\Pi_1^{\dagger\beta}\gamma_{\delta}(\slash{p}_c-\slash{k}_1-\slash{q}+m_c)
\gamma_{\nu}\Bigl(\slash{p}_c+(\slash{k}_2-\slash{k}_1)-\slash{q}+m_c\Bigr)
\gamma_{\sigma}\Bigl(\slash{p}_c+(\slash{k}_2-\slash{k}_1)+m_c\Bigr)\slash{p}]
\nonumber\\
&&\times{1\over [p_c+(k_2-k_1)-q]^2-m_c^2}
{1\over [(p_c-k_1-q)^2-m_c^2]^2}{1\over (p_c-q)^2-m_c^2}
\Pi_{\alpha\beta}g^{\tau\delta}_{\perp}(k_1+q-2p_c)
\nonumber\\
&&\delta\Bigl((k_1+q-2p_c)^2\Bigr),
\eeq
where $p_c$ and $m_c$ are respectively the momentum and the mass of the charm quark
and $g^{\tau\delta}_{\perp}(k_{1,2}+q-2p_c)$ is the polarization tensor of the unobserved gluon. 
The color-singlet hadronization gives $\Pi_{\alpha\beta}$ and $\Pi_1^{\alpha}$  which are defined as
\beq
\Pi_{\alpha\beta}=-g_{\alpha\beta}+{P_{h\alpha}P_{h\beta}\over P_h^2},\hspace{5mm}
\Pi^{\alpha}_1={1\over \sqrt{8m_c^2}}(\slash{p}_c-m_c)\gamma^{\alpha}(\slash{p}_c+m_c).
\eeq
The important difference from the pion production is that the trace factors ${\rm Tr}[\cdots]$ are
separately closed 
in the amplitude level. When the coherent gluon line with momentum $k_2-k_1$
is attached to different closed charm quark loops,
the two diagrams in FIG. 4 are topologically different from each other
and thus they are not canceled at $x_1=x_2$ in the collinear limit. 
As a result, some terms from $k_{1,2}$-derivatives are not canceled,  which means that 
the condition (\ref{additional}) is not satisfied within NRQCD framework. In the next section, we will show that the new calculation technique does not rely on the condition 
(\ref{additional}) and thus it can be applied to the heavy quarkonium production.


\section{New approach to the twist-3 gluon Sivers effect}

In this section, we show the new calculation technique for the twist-3 gluon Sivers effect. We never use 
the identity (\ref{pole_separation}), which is the main difference from the conventional pole 
calculation.
\begin{figure}[h]
\begin{center}
  \includegraphics[height=6cm,width=12cm]{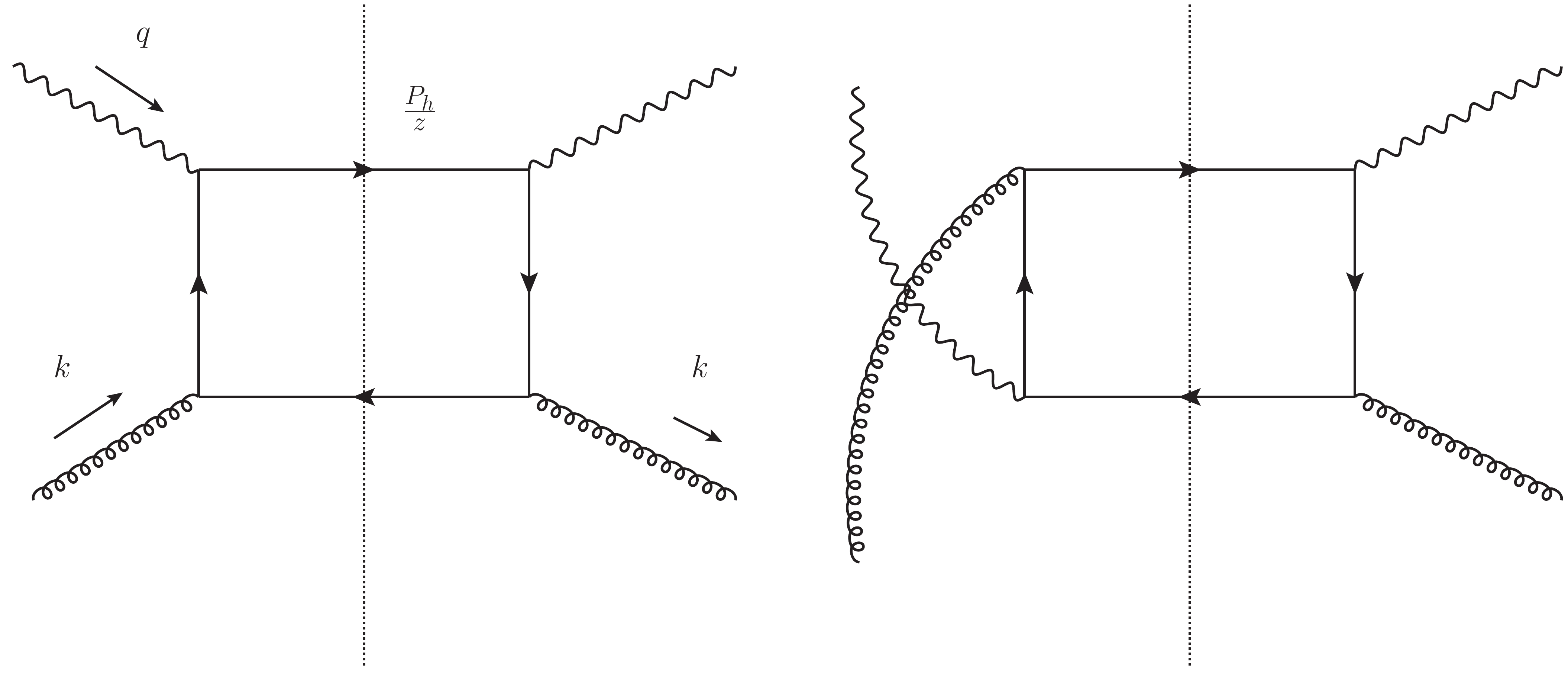}

  \includegraphics[height=6cm,width=12cm]{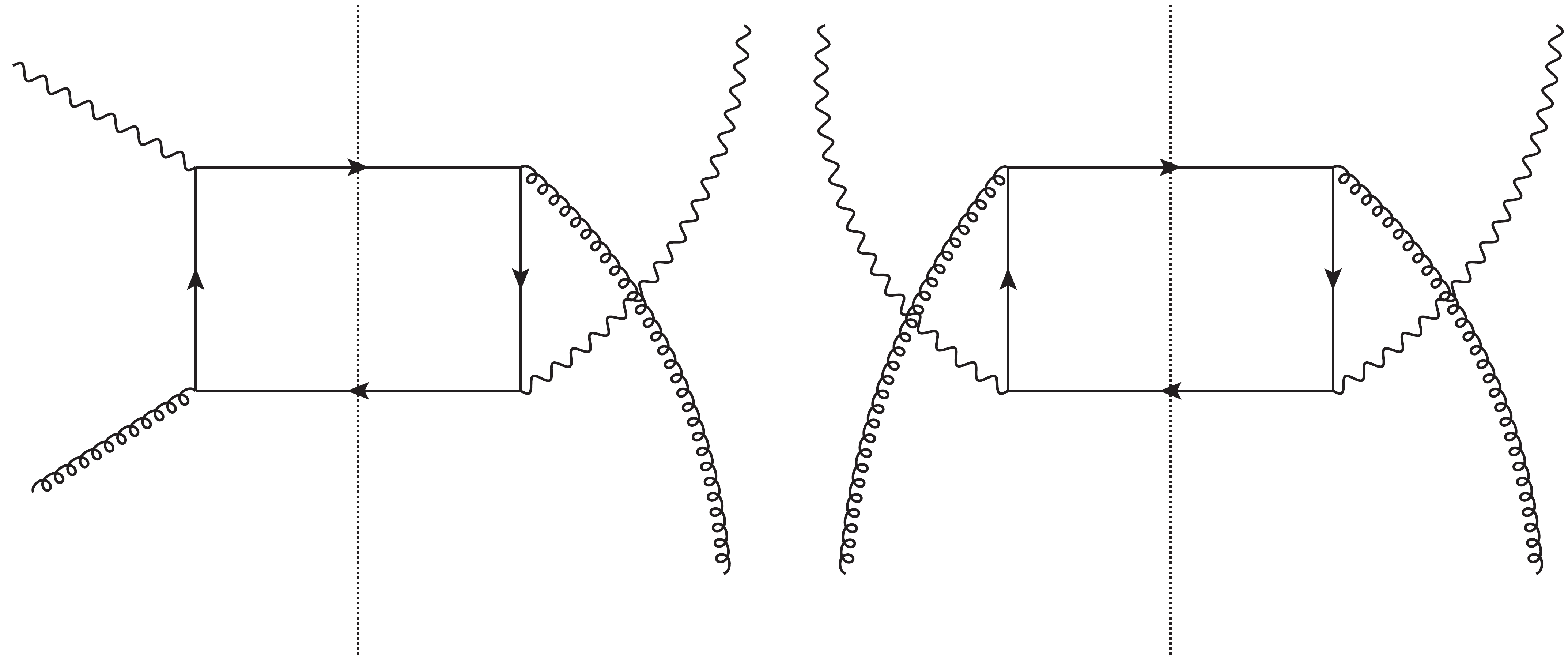}
\end{center}
 \caption{$S^{ab}_{\mu\nu}(k)$ is given by the sum of these diagrams.}
\end{figure}
\begin{figure}[h]
\begin{center}
  \includegraphics[height=6cm,width=12cm]{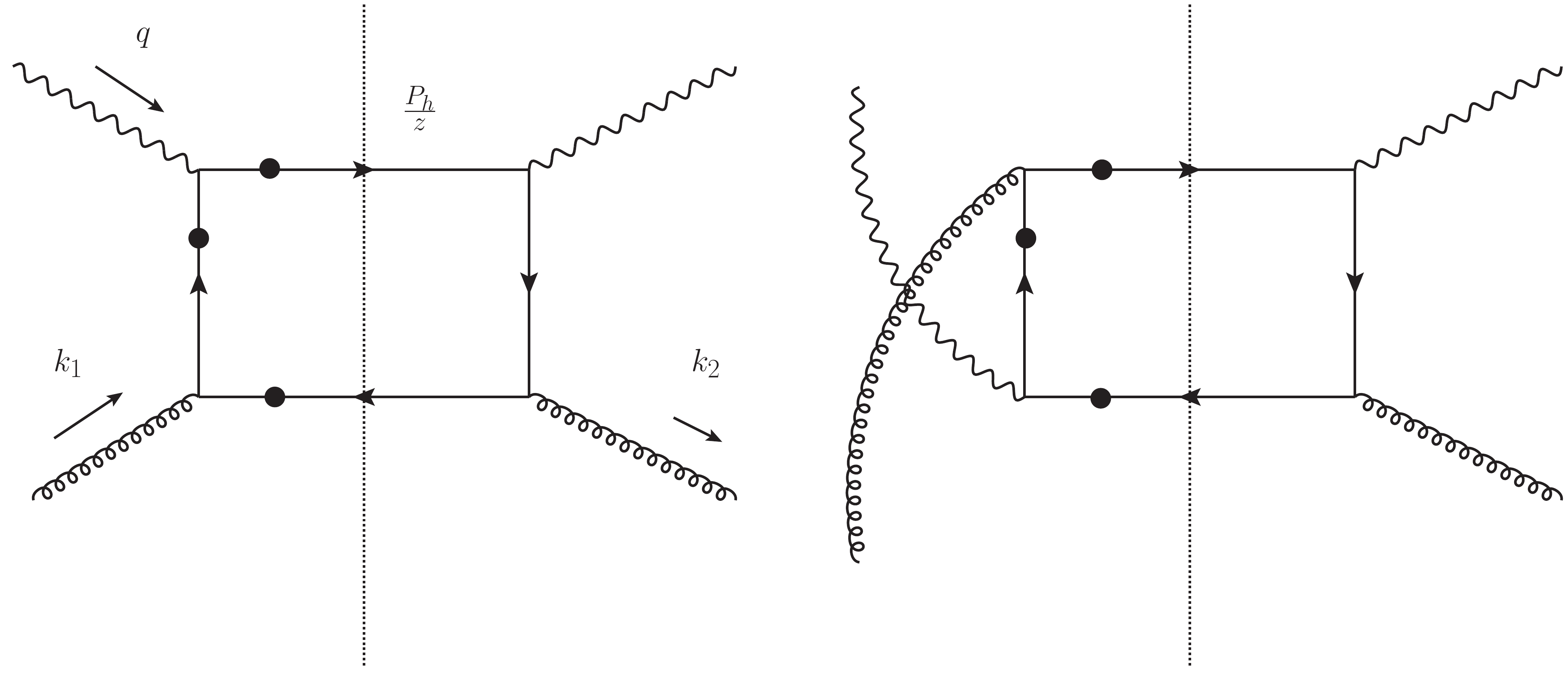}

  \includegraphics[height=6cm,width=12cm]{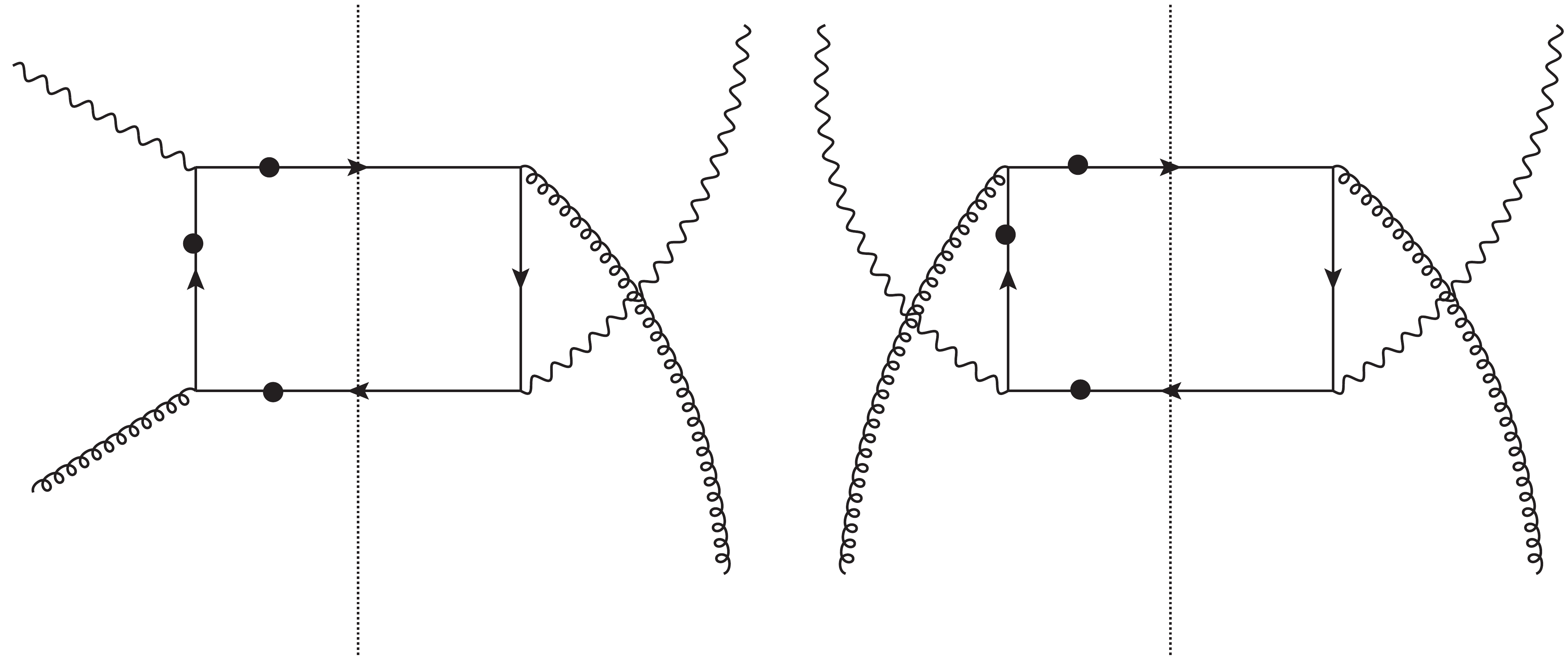}
\end{center}
 \caption{$S^{abc}_{\mu\nu\lambda}(k_1,k_2)$ is given by the sum of these diagrams and their complex
conjugate diagrams. The coherent gluon line with the momentum $k_2-k_1$ is connected
to each black dot and thus there are 12 diagrams in this figure.}
\end{figure}
We take into account the normal $2\to 2$-scattering hard part $S^{ab}_{\mu\nu}(k)$ in FIG. 5
without the coherent gluon because we do not separate the real and the imaginary contributions at the 
beginning,
\beq
w(p,q,{P_h\over z})&=&
\int{d^4k\over (2\pi)^4}
\int{d^4\xi}\,e^{ik\cdot \xi}
\la pS_{\perp}|A_b^{\nu}(0)A_a^{\mu}(\xi)|pS_{\perp}\ra
S^{ab}_{\mu\nu}(k)
\nonumber\\
&&+{1\over 2}\int{d^4k_1\over (2\pi)^4}\int{d^4k_2\over (2\pi)^4}
\int{d^4\xi}\int{d^4\eta}\,e^{ik_1\cdot \xi}e^{i\eta\cdot(k_2-k_1)}
\la pS_{\perp}|A_b^{\nu}(0)gA_c^{\lambda}(\eta)A_a^{\mu}(\xi)|pS_{\perp}\ra
S^{abc}_{\mu\nu\lambda}(k_1,k_2),
\label{w_nonpole}
\eeq
where the factor $1/2$ in the second term in the RHS is needed from the interchange symmetry 
of the external gluon lines. 
The hard part $S^{abc}_{\mu\nu\lambda}(k_1,k_2)$ is given by the sum of diagrams in FIG. 6.
We can show WTIs for hard parts $S^{ab}_{\mu\nu}(k)$ and $S^{abc}_{\mu\nu\lambda}(k_1,k_2)$ \cite{Hatta:2013wsa} as
\beq
k^{\mu}S_{\mu\nu}(k)&=&k^{\nu}S_{\mu\nu}(k)=0,
\label{WTI4}
\\
(k_2-k_1)^{\lambda}S^{abc}_{\mu\nu\lambda}(k_1,k_2)&=&
if^{abc}\Bigl(S_{\mu\nu}(k_2)-S_{\mu\nu}(k_1)\Bigr),
\label{WTI5}
\\
k_1^{\mu}S^{abc}_{\mu\nu\lambda}(k_1,k_2)&=&-if^{abc}S_{\lambda\nu}(k_2),
\label{WTI6}
\\
k_2^{\nu}S^{abc}_{\mu\nu\lambda}(k_1,k_2)&=&-if^{abc}S_{\mu\lambda}(k_1),
\label{WTI7}
\eeq
where we used the color averaged hard part
$S_{\mu\nu}(k)={1\over N_c^2-1}\delta^{ab}S_{\mu\nu}^{ab}(k)$.
Note that WTIs (\ref{WTI5}) - (\ref{WTI7}) have additional terms in the RHS
compared to (\ref{WTI1}) - (\ref{WTI3})
because the delta function from the pole contribution does not exist in the present 
case. We summarize the whole derivation of the gauge-invariant twist-3 contribution in Appendix A
because it is technically complicated and lengthy.
The twist-3 contribution from $w_{\rho\sigma}(p,q,{P_h\over z})$ is given in terms of 
gauge-invariant matrix elements,
\beq
w_{\rho\sigma}(p,q,{P_h\over z})&=&
\omega^{\mu}_{\ \alpha}\omega^{\nu}_{\ \beta}\int{dx\over x^2}
\Phi^{\alpha\beta}(x)S_{\mu\nu,\rho\sigma}(xp)
+\omega^{\mu}_{\ \alpha}\omega^{\nu}_{\ \beta}\omega^{\lambda}_{\ \gamma}
\int{dx\over x^2}\Phi^{\alpha\beta\gamma}_{\partial}(x)
{\partial\over \partial k^{\lambda}}S_{\mu\nu,\rho\sigma}(k)\Bigr|_{k=xp}
\nonumber\\
&&-{1\over 2}\omega^{\mu}_{\ \alpha}\omega^{\nu}_{\ \beta}\omega^{\lambda}_{\ \gamma}
\int dx_1\int dx_2 \Bigl({-if^{abc}\over N_c(N_c^2-1)}N^{\alpha\beta\gamma}(x_1,x_2)
+{N_cd^{abc}\over (N_c^2-4)(N_c^2-1)}O^{\alpha\beta\gamma}(x_1,x_2)\Bigr)
\nonumber\\
&&\times {1\over x_1-i\epsilon}{1\over x_2+i\epsilon}
{1\over x_2-x_1-i\epsilon}S^{abc}_{\mu\nu\lambda,\rho\sigma}(x_1p,x_2p),
\label{w_final}
\eeq
where we restored the virtual photon indices $\rho,\sigma$.
We would like to emphasize that we only used WTIs (\ref{WTI4})-(\ref{WTI7}) in the derivation
and no additional conditions like ({\ref{additional}}) were needed.
The contribution from the matrix element
\beq
\Phi^{\alpha\beta}(x)=\int{d\lambda\over 2\pi}e^{i\lambda x}
\la pS_{\perp}|F^{\beta n}
(0)F^{\alpha n}(\lambda n)|pS_{\perp}\ra,
\eeq
is canceled in the SSA due to the $PT$-invariance.
The hard part $S^{abc}_{\mu\nu\lambda,\rho\sigma}(x_1p,x_2p)$ can be separated
into two parts,
\beq
S^{abc}_{\mu\nu\lambda,\rho\sigma}(x_1p,x_2p)
=H^{L\,abc}_{\mu\nu\lambda,\rho\sigma}(x_1p,x_2p)2\pi\delta\Bigl((x_2p+q-{P_h\over z})^2\Bigr)
+H^{R\,abc}_{\mu\nu\lambda,\rho\sigma}(x_1p,x_2p)2\pi\delta\Bigl((x_1p+q-{P_h\over z})^2\Bigr),
\eeq
where the superscript $L(R)$ means that the coherent gluon is connected to the left(right) side of 
the cut. We can check that each hard part has the following pole structure(See discussion around
Eq. (47) in \cite{Xing:2019ovj}).
\beq
H^{L\,abc}_{\mu\nu\lambda,\rho\sigma}(x_1p,x_2p)=
H^{L0\,abc}_{\mu\nu\lambda,\rho\sigma}(x_1p,x_2p)
+{x_2\over x_1-i\epsilon}H^{L1\,abc}_{\mu\nu\lambda,\rho\sigma}(x_1p,x_2p)
+{x_2\over x_2-x_1-i\epsilon}H^{L2\,abc}_{\mu\nu\lambda,\rho\sigma}(x_1p,x_2p),
\eeq
\beq
H^{R\,abc}_{\mu\nu\lambda,\rho\sigma}(x_1p,x_2p)=
H^{R0\,abc}_{\mu\nu\lambda,\rho\sigma}(x_1p,x_2p)
+{x_1\over x_2+i\epsilon}H^{R1\,abc}_{\mu\nu\lambda,\rho\sigma}(x_1p,x_2p)
+{x_1\over x_2-x_1-i\epsilon}H^{R2\,abc}_{\mu\nu\lambda,\rho\sigma}(x_1p,x_2p).
\eeq
Substituting (\ref{kinematical}), (\ref{C-even}) and (\ref{C-odd}) into (\ref{w_final}), we can show the 
general form of the twist-3 contribution from $w_{\rho\sigma}(p,q,{P_h\over z})$,
\beq
w_{\rho\sigma}(p,q,{P_h\over z})&=&2\pi
\int{dx\over x^2}\delta\Bigl((xp+q-{P_h\over z})^2\Bigr)
\nonumber\\
&&\Biggl[\Bigl(x{d\over dx}G_T^{(1)}(x)-2{G_T^{(1)}(x)}\Bigr)
H^{G1}_{\rho\sigma}+{G_T^{(1)}(x)}H^{G2}_{\rho\sigma}
+\Bigl(x{d\over dx}\Delta H_T^{(1)}(x)-2{\Delta H_T^{(1)}(x)}\Bigr)
H^{H1}_{\rho\sigma}+{\Delta H_T^{(1)}(x)}H^{H2}_{\rho\sigma}
\nonumber\\
&&+\int dx'\sum_i\Bigl[\Bigl({1\over x'-i\epsilon}H^{Ni}_{1L\rho\sigma}
+{1\over x-x'-i\epsilon}H^{Ni}_{2L\rho\sigma}+{x\over (x'-i\epsilon)^2}
H^{Ni}_{3L\rho\sigma}+{x\over (x-x'-i\epsilon)^2}H^{Ni}_{4L\rho\sigma}
\Bigr)N^i(x',x)
\nonumber\\
&&+\Bigl({1\over x'+i\epsilon}H^{Ni}_{1R\rho\sigma}
+{1\over x-x'+i\epsilon}H^{Ni}_{2R\rho\sigma}+{x\over (x'+i\epsilon)^2}
H^{Ni}_{3R\rho\sigma}+{x\over (x-x'+i\epsilon)^2}H^{Ni}_{4R\rho\sigma}
\Bigr)N^i(x,x')
\nonumber\\
&&+\Bigl({1\over x'-i\epsilon}H^{Oi}_{1L\rho\sigma}
+{1\over x-x'-i\epsilon}H^{Oi}_{2L\rho\sigma}+{x\over (x'-i\epsilon)^2}
H^{Oi}_{3L\rho\sigma}+{x\over (x-x'-i\epsilon)^2}H^{Oi}_{4L\rho\sigma}
\Bigr)O^i(x',x)
\nonumber\\
&&+\Bigl({1\over x'+i\epsilon}H^{Oi}_{1R\rho\sigma}
+{1\over x-x'+i\epsilon}H^{Oi}_{2R\rho\sigma}+{x\over (x'+i\epsilon)^2}
H^{Oi}_{3R\rho\sigma}+{x\over (x-x'+i\epsilon)^2}H^{Oi}_{4R\rho\sigma}
\Bigr)O^i(x,x')
\Bigr]
\Biggr],
\label{kinematical_dynamical}
\eeq
where we used shorthand notations
\beq
N^{1,2,3}(x',x)=\{N(x',x),N(x,x-x'),N(x',x'-x)\},\hspace{5mm}O^{1,2,3}(x',x)=\{O(x',x),O(x,x-x'),O(x',x'-x)\}.
\hspace{2mm}
\eeq
Furthermore, the dynamical part can be simplified by interchanging the integral variable as
$x'\to x-x'$ and using the symmetries (\ref{symmetries}),
\beq
&&\int dx'\sum_i\Bigl[\Bigl(
{1\over x-x'-i\epsilon}H^{Ni}_{SL\rho\sigma}
+{x\over (x-x'-i\epsilon)^2}H^{Ni}_{DL\rho\sigma}
\Bigr)N^i(x',x)
\nonumber\\
&&+\Bigl({1\over x-x'+i\epsilon}H^{Ni}_{SR\rho\sigma}
+{x\over (x-x'+i\epsilon)^2}H^{Ni}_{DR\rho\sigma}
\Bigr)N^i(x,x')
\nonumber\\
&&+\Bigl({1\over x-x'-i\epsilon}H^{Oi}_{SL\rho\sigma}
+{x\over (x-x'-i\epsilon)^2}H^{Oi}_{DL\rho\sigma}
\Bigr)O^i(x',x)
\nonumber\\
&&+\Bigl({1\over x-x'+i\epsilon}H^{Oi}_{SR\rho\sigma}
+{x\over (x-x'+i\epsilon)^2}H^{Oi}_{DR\rho\sigma}
\Bigr)O^i(x,x').
\eeq
The hard parts were recombined as
\beq
H^{N1(O1)}_{SL\rho\sigma}&=&H^{N1(O1)}_{2L\rho\sigma}+H^{N2(O2)}_{1L\rho\sigma},\hspace{5mm}
H^{N1(O1)}_{DL\rho\sigma}=H^{N1(O1)}_{4L\rho\sigma}+H^{N2(O2)}_{3L\rho\sigma},
\nonumber\\
H^{N2(O2)}_{SL\rho\sigma}&=&H^{N2(O2)}_{2L\rho\sigma}+H^{N1(O1)}_{1L\rho\sigma},\hspace{5mm}
H^{N2(O2)}_{DL\rho\sigma}=H^{N2(O2)}_{4L\rho\sigma}+H^{N1(O1)}_{3L\rho\sigma},
\nonumber\\
H^{N3(O3)}_{SL\rho\sigma}&=&H^{N3(O3)}_{2L\rho\sigma}\mp H^{N3(O3)}_{1L\rho\sigma},\hspace{5mm}
H^{N3(O3)}_{DL\rho\sigma}=H^{N3(O3)}_{4L\rho\sigma}\mp H^{N3(O3)}_{3L\rho\sigma},
\nonumber\\
H^{N1(O1)}_{SR\rho\sigma}&=&H^{N1(O1)}_{2R\rho\sigma}+H^{N3(O3)}_{1R\rho\sigma},\hspace{5mm}
H^{N1(O1)}_{DR\rho\sigma}=H^{N1(O1)}_{4R\rho\sigma}+H^{N3(O3)}_{3R\rho\sigma},
\nonumber\\
H^{N2(O2)}_{SR\rho\sigma}&=&H^{N2(O2)}_{2R\rho\sigma}\mp H^{N2(O2)}_{1R\rho\sigma},\hspace{5mm}
H^{N2(O2)}_{DR\rho\sigma}=H^{N2(O2)}_{4R\rho\sigma}\mp H^{N2(O2)}_{3R\rho\sigma},
\nonumber\\
H^{N3(O3)}_{SR\rho\sigma}&=&H^{N3(O3)}_{2R\rho\sigma}+H^{N1(O1)}_{1R\rho\sigma},\hspace{5mm}
H^{N3(O3)}_{DR\rho\sigma}=H^{N3(O3)}_{4R\rho\sigma}+H^{N1(O1)}_{3R\rho\sigma}.
\eeq
The cross section formula is derived by calculating $L^{\rho\sigma}w_{\rho\sigma}(p,q,{P_h\over z})$.
The lepton tensor $L^{\rho\sigma}$ is conventionally expended as
\beq
L^{\rho\sigma}=\sum_{k=1}^9(L^{\mu\nu}{\cal V}_{k\,\mu\nu})\tilde{\cal V}^{\rho\sigma}_k,
\eeq
where tensors ${\cal V}_{k\,\mu\nu}$ and $\tilde{\cal V}^{\mu\nu}_k$ are given in 
\cite{Beppu:2010qn,Meng:1991da}. 
Symmetric tensors $k=1,2,3,4,8,9$ give nonzero contributions to the cross section.
From a direct calculation, we have observed
\beq
&&\tilde{\cal V}_k^{\rho\sigma}H^{N1(O1)}_{SL\rho\sigma}=-\tilde{\cal V}_k^{\rho\sigma}
H^{N1(O1)}_{SR\rho\sigma},\hspace{3mm}
\tilde{\cal V}_k^{\rho\sigma}H^{N2(O2)}_{SL\rho\sigma}=-\tilde{\cal V}_k^{\rho\sigma}
H^{N3(O3)}_{SR\rho\sigma},\hspace{3mm}
\tilde{\cal V}_k^{\rho\sigma}H^{N3(O3)}_{SL\rho\sigma}=-\tilde{\cal V}_k^{\rho\sigma}
H^{N2(O2)}_{SR\rho\sigma},
\nonumber\\
&&\tilde{\cal V}_k^{\rho\sigma}H^{N1(O1)}_{DL\rho\sigma}=-\tilde{\cal V}_k^{\rho\sigma}
H^{N1(O1)}_{DR\rho\sigma},\hspace{3mm}
\tilde{\cal V}_k^{\rho\sigma}H^{N2(O2)}_{DL\rho\sigma}=-\tilde{\cal V}_k^{\rho\sigma}
H^{N2(O2)}_{DR\rho\sigma}=
\tilde{\cal V}_k^{\rho\sigma}H^{N3(O3)}_{DL\rho\sigma}=-\tilde{\cal V}_k^{\rho\sigma}
H^{N3(O3)}_{DR\rho\sigma},
\eeq
for any $k$.
Thus we can perform contour integrations as
\beq
&&{1\over 2}\int dx'\Bigl({1\over x-x'-i\epsilon}-{1\over x-x'+i\epsilon}\Bigr)
\Bigl(N(x',x)+N(x,x')\Bigr)=2\pi i N(x,x),
\\
&&\int dx'\Bigl({1\over x-x'-i\epsilon}-{1\over x-x'+i\epsilon}\Bigr)
N(x,x-x')=2\pi i N(x,0),
\\
&&\int dx'\Bigl({1\over x-x'-i\epsilon}-{1\over x-x'+i\epsilon}\Bigr)
N(x',x'-x)=2\pi i N(x,0),
\\
&&{1\over 2}\int dx'\Bigl({1\over (x-x'-i\epsilon)^2}-{1\over (x-x'+i\epsilon)^2}\Bigr)
\Bigl(N(x',x)+N(x,x')\Bigr)=-\pi i {d\over dx}N(x,x),
\\
&&\int dx'\Bigl({1\over (x-x'-i\epsilon)^2}-{1\over (x-x'+i\epsilon)^2}\Bigr)
\Bigl(N(x,x-x')+N(x',x'-x)\Bigr)=-2\pi i {d\over dx}N(x,0),
\eeq
and the same results for $O^i(x',x)$. The kinematical functions in (\ref{kinematical_dynamical}) 
can be eliminated by using
(\ref{relations}). Then we can finally express the cross section formula only in terms of 
$O(x,x), O(x,0), N(x,x)$ and $N(x,0)$ as shown by the pole calculation \cite{Beppu:2010qn},
\beq
\frac{d^{6}\Delta\sigma}{dx_{bj}dQ^{2}dz_{f}dP^{2}_{h}d\phi d\chi}&=&
{\alpha^2_{em}\alpha_s\over 16\pi^2 z_{f}x_{bj}^2S^2_{ep}}(-{\pi M_N\over 2})\sum_a e^2_a
\sum_k{\cal A}_k(\phi-\chi){\cal S}_k(\Phi_s-\chi)\int{{dz\over z^2}}D_a(z)
\int{dx\over x}\delta\Bigl((xp+q-{P_h\over z})^2\Bigr)
\nonumber\\
&&\Biggl[\Bigl({d\over dx}O(x,x)+{d\over dx}N(x,x)\Bigr)\Delta\hat{\sigma}^1_k
+\Bigl({O(x,x)\over x}+{N(x,x)\over x}\Bigr)\Bigl(\Delta\hat{\sigma}^3_k-2\Delta\hat{\sigma}^1_k\Bigr)
\nonumber\\
&&+\Bigl({d\over dx}O(x,0)-{d\over dx}N(x,0)\Bigr)\Delta\hat{\sigma}^2_k
+\Bigl({O(x,0)\over x}-{N(x,0)\over x}\Bigr)\Bigl(\Delta\hat{\sigma}^4_k-2\Delta\hat{\sigma}^2_k\Bigr)
\Biggr],
\eeq
where forms of ${\cal A}_k(\phi)$, ${\cal S}_k(\Phi_s)$ are given in \cite{Beppu:2010qn,Koike:2011ns} 
and the index $a=u,d,s,\cdots$
denotes the flavor of the quark fragmenting into the pion. 
We have confirmed that all hard cross sections for $k=1,2,3,4,8,9$ are consistent with those in 
\cite{Beppu:2010qn} in the massless limit
$m_c=0$. Our calculation shows that the condition (\ref{additional}) in the conventional pole calculation is 
an unnecessary condition caused by the separation of the pole part in (\ref{pole_separation}). 
All necessary conditions are only WTIs (\ref{WTI4})-(\ref{WTI7}) followed by
the gauge invariant structure in the amplitude level. The gauge invariance is the fundamental property of
a scattering amplitude and it should be satisfied in reasonable perturbative QCD frameworks including NRQCD.
Our new method extends the applicability of the twist-3 technique for the gluon Sivers effect, which is of great importance in the analysis of SSAs in heavy quarkonium productions. 

The unknown nonperturbative functions $O(x,x)$, $O(x,0)$, $N(x,x)$ and $N(x,0)$ can 
be determined through the standard global analysis of data. 
The cross section formula for the SSA has been derived in some processes($D$-meson production in SIDIS\cite{Kang:2008qh,Beppu:2010qn}
 and $pp$\cite{Kang:2008ih,Koike:2011mb}
, direct photon production 
and Drell-Yan in $pp$\cite{Koike:2011nx})
and the experimental investigation has just begun\cite{PHENIX:2021irw,PHENIX:2022znm}.
We expect that the $C$-odd contribution from $O(x,x)$ and $O(x,0)$ is canceled in the quarkonium
production and thus the SSAs in heavy quarkonium productions could play a role in the 
independent determination of
$C$-even functions $N(x,x)$ and $N(x,0)$.


\section{Summary}

In this paper, we have proposed the new calculation method for the twist-3 gluon Sivers effect 
in the collinear factorization approach and confirmed that known results calculated by the conventional 
pole method can be successfully reproduced. 
Our calculation has clarified that the pole calculation was using 
an unnecessary condition which hinders the application of the twist-3 framework to heavy 
quarkonium productions. Our new method just requires basic WTIs which are satisfied in 
reasonable perturbative QCD frameworks including NRQCD for the heavy quarkonium production.
SSAs in heavy meson productions are ideal observables for the study of the gluon Sivers 
effect. The future EIC experiment is expected to measure those SSAs 
in the high transverse momentum region of produced hadrons as the RHIC experiment has done 
in the past couple of decades.
Our method gives a theoretical basis to the analysis of SSAs observed in a wide range of 
the transverse momentum.


\appendix

\section{Derivation of Eq. (\ref{w_final})}

We decompose a momentum vector into the longitudinal part and the other part as
\beq
k^{\mu}=(k\cdot n)p^{\mu}+\omega^{\mu}_{\ \rho}k^{\rho},
\eeq
in order to extract twist-3 contributions from $w(p,q,{P_h\over z})$.
Consequently the WTI (\ref{WTI4}) is rewritten as
\beq
k^{\mu}S_{\mu\nu}(k)=(k\cdot n)S_{p\nu}(k)+\omega^{\mu}_{\ \rho}k^{\rho}S_{\mu\nu}(k)=0
,\hspace{5mm}
k^{\nu}S_{\mu\nu}(k)=(k\cdot n)S_{\mu p}(k)+\omega^{\nu}_{\ \sigma}k^{\sigma}S_{\mu\nu}(k)=
0.
\label{WTI_identity}
\eeq
We can calculate the first term in (\ref{w_nonpole}) as
\beq
&&\int{d^4k\over (2\pi)^4}
\int{d^4\xi}\,e^{ik\cdot \xi}
\la pS_{\perp}|A_a^{\sigma}(0)A_a^{\rho}(\xi)|pS_{\perp}\ra
g^{\mu}_{\ \rho}g^{\nu}_{\ \sigma}S_{\mu\nu}(k)
\nonumber\\
&=&\int{d^4k\over (2\pi)^4}
\int{d^4\xi}\,e^{ik\cdot \xi}
\la pS_{\perp}|A_a^{\sigma}(0)A_a^{\rho}(\xi)|pS_{\perp}\ra
(p^{\mu}n_{\rho}+\omega^{\mu}_{\ \rho})(p^{\nu}n_{\sigma}+\omega^{\nu}_{\ \sigma})S_{\mu\nu}(k)
\nonumber\\
&=&\int{d^4k\over (2\pi)^4}{1\over (k\cdot n)^2}\omega^{\mu}_{\ \tau}\omega^{\nu}_{\ \delta}
\int{d^4\xi}\,e^{ik\cdot \xi}
\la pS_{\perp}|A_a^{\sigma}(0)A_a^{\rho}(\xi)|pS_{\perp}\ra
(k^{\tau}n_{\rho}-(k\cdot n)g^{\tau}_{\ \rho})
(k^{\delta}n_{\sigma}-(k\cdot n)g^{\delta}_{\ \sigma})S_{\mu\nu}(k)
\nonumber\\
&=&\int{d^4k\over (2\pi)^4}{1\over (k\cdot n)^2}\omega^{\mu}_{\ \tau}\omega^{\nu}_{\ \delta}
\int{d^4\xi}\,e^{ik\cdot \xi}
\la pS_{\perp}|F_a^{(0)\delta n}(0)F_a^{(0)\tau n}(\xi)|pS_{\perp}\ra
S_{\mu\nu}(k),
\eeq
where $F_a^{(0)\tau n}(\xi)\equiv \partial^{\tau}A_a^{n}(\xi)-\partial^nA_a^{\tau}(\xi)$
and we used (\ref{WTI_identity}) in the second equality.
We perform the collinear expansion for the hard part,
\beq
S_{\mu\nu}(k)\simeq S_{\mu\nu}((k\cdot n)p)+{\partial\over \partial k^{\lambda}}
S_{\mu\nu}(k)\Bigr|_{k=(k\cdot n)p}\omega^{\lambda}_{\ \kappa}k^{\kappa}.
\label{collinear_twist-2}
\eeq
As a result, we obtain the twist-3 contribution from the first term in (\ref{w_nonpole}),
\beq
&&\int{d^4k\over (2\pi)^4}{1\over (k\cdot n)^2}\omega^{\mu}_{\ \tau}\omega^{\nu}_{\ \delta}
\int{d^4\xi}\,e^{ik\cdot \xi}
\la pS_{\perp}|F_a^{(0)\delta n}(0)F_a^{(0)\tau n}(\xi)|pS_{\perp}\ra
\Bigl(S_{\mu\nu}((k\cdot n)p)+{\partial\over \partial k^{\lambda}}
S_{\mu\nu}(k)\Bigr|_{k=(k\cdot n)p}\omega^{\lambda}_{\ \kappa}k^{\kappa}\Bigr)
\nonumber\\
&=&\omega^{\mu}_{\ \tau}\omega^{\nu}_{\ \delta}
\int{dx\over x^2}\int{d\lambda\over 2\pi}e^{i\lambda x}
\la pS_{\perp}|F_a^{(0)\delta n}(0)F_a^{(0)\tau n}(\lambda n)|pS_{\perp}\ra
S_{\mu\nu}(xp)
\nonumber\\
&&+i\omega^{\mu}_{\ \tau}\omega^{\nu}_{\ \delta}\omega^{\lambda}_{\ \kappa}
\int{dx\over x^2}\int{d\lambda\over 2\pi}e^{i\lambda x}
\la pS_{\perp}|F_a^{(0)\delta n}(0)\partial^{\kappa}F_a^{(0)\tau n}(\lambda n)|pS_{\perp}\ra
{\partial\over \partial k^{\lambda}}S_{\mu\nu}(k)\Bigr|_{k=xp}.
\eeq
We next calculate the second term in (\ref{w_nonpole}). We rewrite (\ref{WTI5})-(\ref{WTI7}) as,
\beq
(k_2-k_1)^{\lambda}S^{abc}_{\mu\nu\lambda}(k_1,k_2)&=&
(k_2\cdot n-k_1\cdot n)S^{abc}_{\mu\nu p}(k_1,k_2)
+\omega^{\lambda}_{\ \phi}(k_2-k_1)^{\phi}S^{abc}_{\mu\nu\lambda}(k_1,k_2)
=if^{abc}\Bigl(S_{\mu\nu}(k_2)-S_{\mu\nu}(k_1)\Bigr),\hspace{3mm}
\\
k_1^{\mu}S^{abc}_{\mu\nu\lambda}(k_1,k_2)&=&
k_1\cdot nS^{abc}_{p \nu\lambda}(k_1,k_2)
+\omega^{\mu}_{\ \rho}k_1^{\rho}S^{abc}_{\mu\nu\lambda}(k_1,k_2)
=-if^{abc}S_{\lambda\nu}(k_2),
\\
k_2^{\nu}S^{abc}_{\mu\nu\lambda}(k_1,k_2)&=&
k_2\cdot nS^{abc}_{\mu p\lambda}(k_1,k_2)
+\omega^{\nu}_{\ \sigma}k_2^{\sigma}S^{abc}_{\mu\nu\lambda}(k_1,k_2)
=-if^{abc}S_{\mu\lambda}(k_1).
\eeq
Then we can show
\beq
&&{1\over 2}\int{d^4k_1\over (2\pi)^4}\int{d^4k_2\over (2\pi)^4}
\int{d^4\xi}\int{d^4\eta}\,e^{ik_1\cdot \xi}e^{i\eta\cdot(k_2-k_1)}
\la pS_{\perp}|A_b^{\sigma}(0)gA_c^{\phi}(\eta)A_a^{\rho}(\xi)|pS_{\perp}\ra
\nonumber\\
&&\times(p^{\mu}n_{\rho}+\omega^{\mu}_{\ \rho})
(p^{\nu}n_{\sigma}+\omega^{\nu}_{\ \sigma})
(p^{\lambda}n_{\phi}+\omega^{\lambda}_{\ \phi})S^{abc}_{\mu\nu\lambda}(k_1,k_2)
\nonumber\\
&=&{1\over 2}\int{d^4k_1\over (2\pi)^4}\int{d^4k_2\over (2\pi)^4}
\int{d^4\xi}\int{d^4\eta}\,e^{ik_1\cdot \xi}e^{i\eta\cdot(k_2-k_1)}
\la pS_{\perp}|A_b^{\sigma}(0)gA_c^{\phi}(\eta)A_a^{\rho}(\xi)|pS_{\perp}\ra
\nonumber\\
&&\times \Bigl[-{1\over k_1\cdot n-i\epsilon}{1\over k_2\cdot n+i\epsilon}
{1\over k_2\cdot n-k_1\cdot n-i\epsilon}
\omega^{\mu}_{\ \tau}\omega^{\nu}_{\ \delta}\omega^{\lambda}_{\ \kappa}
(k_1^{\tau}n_{\rho}-k_1\cdot ng^{\tau}_{\ \rho})
(k_2^{\delta}n_{\sigma}-k_2\cdot ng^{\delta}_{\ \sigma})
\nonumber\\
&&\times
((k_2-k_1)^{\kappa}n_{\phi}-(k_2\cdot n-k_1\cdot n)g^{\kappa}_{\ \phi})S^{abc}_{\mu\nu\lambda}(k_1,k_2)
\nonumber\\
&&+{-if^{abc}\over k_1\cdot n-i\epsilon}n_{\rho}{1\over (k_2\cdot n+i\epsilon)^2}
\omega^{\nu}_{\ \delta}\omega^{\lambda}_{\ \kappa}(k_2^{\delta}n_{\sigma}-(k_2\cdot n)
g^{\delta}_{\ \sigma})
(k_2^{\kappa}n_{\phi}-(k_2\cdot n)g^{\kappa}_{\ \phi})S_{\lambda\nu}(k_2)
\nonumber\\
&&+{-if^{abc}\over k_2\cdot n+i\epsilon}n_{\sigma}{1\over (k_1\cdot n-i\epsilon)^2}
\omega^{\mu}_{\ \tau}\omega^{\lambda}_{\ \kappa}(k_1^{\tau}n_{\rho}-(k_1\cdot n)g^{\tau}_{\ \rho})
(k_1^{\kappa}n_{\phi}-(k_1\cdot n)g^{\kappa}_{\ \phi})S_{\mu\lambda}(k_1)
\nonumber\\
&&+{if^{abc}\over k_2\cdot n-k_1\cdot n-i\epsilon}n_{\phi}{1\over k_1\cdot n-i\epsilon}
{1\over k_2\cdot n+i\epsilon}
\omega^{\mu}_{\ \tau}\omega^{\nu}_{\ \delta}(k_1^{\tau}n_{\rho}-(k_1\cdot n)g^{\tau}_{\ \rho})
(k_2^{\delta}n_{\sigma}-(k_2\cdot n)g^{\delta}_{\ \sigma})S_{\mu\nu}(k_2)
\nonumber\\
&&-{if^{abc}\over k_2\cdot n-k_1\cdot n-i\epsilon}n_{\phi}{1\over k_1\cdot n-i\epsilon}
{1\over k_2\cdot n+i\epsilon}
\omega^{\mu}_{\ \tau}\omega^{\nu}_{\ \delta}(k_1^{\tau}n_{\rho}-(k_1\cdot n)g^{\tau}_{\ \rho})
(k_2^{\delta}n_{\sigma}-(k_2\cdot n)g^{\delta}_{\ \sigma})S_{\mu\nu}(k_1)
\Bigr].
\label{collinear_twist-3}
\eeq
We keep $i\epsilon$-terms in order to correctly perform contour integrations. The signs of 
$i\epsilon$ are uniquely determined from the fact that the diagrams in FIG. 6 have only the final state 
interaction. 
We perform the collinear expansion for $S^{abc}_{\mu\nu\lambda}(k_1,k_2)$,
\beq
S^{abc}_{\mu\nu\lambda}(k_1,k_2)\simeq S^{abc}_{\mu\nu\lambda}((k_1\cdot n)p,(k_2\cdot n)p).
\eeq
The first term in the bracket $[\cdots]$ in (\ref{collinear_twist-3}) can be calculated as
\beq
&&{1\over 2}\int{d^4k_1\over (2\pi)^4}\int{d^4k_2\over (2\pi)^4}
\int{d^4\xi}\int{d^4\eta}\,e^{ik_1\cdot \xi}e^{i\eta\cdot(k_2-k_1)}
\la pS_{\perp}|A_b^{\sigma}(0)gA_c^{\phi}(\eta)A_a^{\rho}(\xi)|pS_{\perp}\ra
\nonumber\\
&&\times \Bigl[-{1\over k_1\cdot n-i\epsilon}{1\over k_2\cdot n+i\epsilon}
{1\over k_2\cdot n-k_1\cdot n-i\epsilon}
\omega^{\mu}_{\ \tau}\omega^{\nu}_{\ \delta}\omega^{\lambda}_{\ \kappa}
(k_1^{\tau}n_{\rho}-k_1\cdot ng^{\tau}_{\ \rho})
(k_2^{\delta}n_{\sigma}-k_2\cdot ng^{\delta}_{\ \sigma})
\nonumber\\
&&\times
((k_2-k_1)^{\kappa}n_{\phi}-(k_2\cdot n-k_1\cdot n)g^{\kappa}_{\ \phi})
S^{abc}_{\mu\nu\lambda}((k_1\cdot n)p,(k_2\cdot n)p)
\Bigr]
\nonumber\\
&=&-{i\over 2}\omega^{\mu}_{\ \tau}\omega^{\nu}_{\ \delta}\omega^{\lambda}_{\ \kappa}
\int dx_1\int dx_2\int{d\lambda\over 2\pi}\int{d\mu\over 2\pi}
e^{i\lambda x_1}e^{i\mu(x_2-x_1)}
\la pS_{\perp}|F_b^{(0)\delta n}(0)gF_c^{(0)\kappa n}(\mu n)F_a^{(0)\tau n}(\lambda n)|pS_{\perp}\ra
\nonumber\\
&&\times {1\over x_1-i\epsilon}{1\over x_2+i\epsilon}
{1\over x_2-x_1-i\epsilon}S^{abc}_{\mu\nu\lambda}(x_1p,x_2p).
\eeq
This term gives $g^1$-term in the dynamical matrix elements (\ref{C-even}) and (\ref{C-odd}).
We use the collinear expansion (\ref{collinear_twist-2}) for other terms in (\ref{collinear_twist-3}). 
We first show 
contributions from the first term of  (\ref{collinear_twist-2}).
The second term in the bracket $[\cdots]$ in (\ref{collinear_twist-3}) can be calculated as 
\beq
&&{1\over 2}\int{d^4k_1\over (2\pi)^4}\int{d^4k_2\over (2\pi)^4}
\int{d^4\xi}\int{d^4\eta}\,e^{ik_1\cdot \xi}e^{i\eta\cdot(k_2-k_1)}
\la pS_{\perp}|A_b^{\sigma}(0)gA_c^{\phi}(\eta)A_a^{\rho}(\xi)|pS_{\perp}\ra
\nonumber\\
&&\times \Bigl[{-if^{abc}\over k_1\cdot n-i\epsilon}n_{\rho}{1\over (k_2\cdot n+i\epsilon)^2}
\omega^{\nu}_{\ \delta}\omega^{\lambda}_{\ \kappa}(k_2^{\delta}n_{\sigma}-(k_2\cdot n)
g^{\delta}_{\ \sigma})
(k_2^{\kappa}n_{\phi}-(k_2\cdot n)g^{\kappa}_{\ \phi})S_{\lambda\nu}((k_2\cdot n)p)\Bigr]
\nonumber\\
&=&-{i\over 2}\int{d^4k_1\over (2\pi)^4}\int{d^4k_2\over (2\pi)^4}
\int{d^4\xi}\int{d^4\eta}\,e^{ik_1\cdot \xi}e^{i\eta\cdot(k_2-k_1)}
{-if^{abc}\over k_1\cdot n-i\epsilon}{1\over (k_2\cdot n+i\epsilon)^2}
\omega^{\mu}_{\ \tau}\omega^{\nu}_{\ \delta}
\la pS_{\perp}|F^{(0)\delta n}_b(0)gA_c^{\phi}(\eta)A_a^{n}(\xi)|pS_{\perp}\ra
\nonumber\\
&&\times (k_2^{\tau}n_{\phi}-(k_2\cdot n)g^{\tau}_{\ \phi})S_{\mu\nu}((k_2\cdot n)p)
\nonumber\\
&=&{1\over 2}\int dx_1\int dx_2\int{d\lambda\over 2\pi}\int{d\mu\over 2\pi}
\,e^{i\lambda x_1}e^{i\mu(x_2-x_1)}
{-if^{abc}\over x_1-i\epsilon}{1\over (x_2+i\epsilon)^2}
\omega^{\mu}_{\ \tau}\omega^{\nu}_{\ \delta}
\nonumber\\
&&\times \Bigl[\la pS_{\perp}|F^{(0)\delta n}_b(0)gF_c^{(0)\tau n}(\mu n)
A_a^{n}(\lambda n)|pS_{\perp}\ra
+\la pS_{\perp}|F^{(0)\delta n}_b(0)gA_c^{n}(\mu n)F_a^{(0)\tau n}(\lambda n)|pS_{\perp}\ra
\nonumber\\
&&+\la pS_{\perp}|F^{(0)\delta n}_b(0)gA_c^{n}(\mu n)
\partial^n A_a^{\tau}(\lambda n)|pS_{\perp}\ra
-\la pS_{\perp}|F^{(0)\delta n}_b(0)gA_c^{\tau}(\mu n)\partial^{n}A_a^{n}(\lambda n)|pS_{\perp}\ra
\Bigr]
S_{\mu\nu}(x_2p)
\nonumber\\
&=&{1\over 2}\int {dx\over x^2}\int{d\lambda\over 2\pi}
\,e^{i\lambda x}
\omega^{\mu}_{\ \tau}\omega^{\nu}_{\ \delta}
\nonumber\\
&&\times \Bigl[-if^{abc}\int^{\infty}_{\lambda}d\mu\la pS_{\perp}|F^{(0)\delta n}_b(0)igA_a^{n}(\mu n)
F_c^{(0)\tau n}(\lambda n)
|pS_{\perp}\ra
-if^{abc}\int^{\infty}_{\lambda}d\mu
\la pS_{\perp}|F^{(0)\delta n}_b(0)igA_c^{n}(\lambda n)F_a^{(0)\tau n}(\mu n)|pS_{\perp}\ra
\nonumber\\
&&+2f^{abc}\la pS_{\perp}|F^{(0)\delta n}_b(0)gA_c^{\tau}(\lambda n)A_a^{n}(\lambda n)|pS_{\perp}\ra
\Bigr]
S_{\mu\nu}(xp),
\label{part1}
\eeq
where we used the integral representation of the theta function,
\beq
\int{dx}{e^{iqx}\over x-i\epsilon}=2\pi i\theta(q).
\eeq
The remaining terms in (\ref{collinear_twist-3}) can be calculated in the same way as
\beq
&&{1\over 2}\int{d^4k_1\over (2\pi)^4}\int{d^4k_2\over (2\pi)^4}
\int{d^4\xi}\int{d^4\eta}\,e^{ik_1\cdot \xi}e^{i\eta\cdot(k_2-k_1)}
\la pS_{\perp}|A_b^{\sigma}(0)gA_c^{\phi}(\eta)A_a^{\rho}(\xi)|pS_{\perp}\ra
\nonumber\\
&&\times \Bigl[
{-if^{abc}\over k_2\cdot n+i\epsilon}n_{\sigma}{1\over (k_1\cdot n-i\epsilon)^2}
\omega^{\mu}_{\ \tau}\omega^{\lambda}_{\ \kappa}(k_1^{\tau}n_{\rho}-(k_1\cdot n)g^{\tau}_{\ \rho})
(k_1^{\kappa}n_{\phi}-(k_1\cdot n)g^{\kappa}_{\ \phi})S_{\mu\lambda}((k_1\cdot n)p)
\Bigr]
\nonumber\\
&=&{1\over 2}\int {dx\over x^2}\int{d\lambda\over 2\pi}
\,e^{i\lambda x}
\omega^{\mu}_{\ \tau}\omega^{\nu}_{\ \delta}
\nonumber\\
&&\times \Bigl[if^{abc}\int^{\infty}_{0}d\mu\la pS_{\perp}|F^{(0)\delta n}_b(\mu n)igA_c^{n}(0)
F_a^{(0)\tau n}(\lambda n)
|pS_{\perp}\ra
+if^{abc}\int^{\infty}_{0}d\mu
\la pS_{\perp}|F^{(0)\delta n}_b(0)igA_a^{n}(\mu n)F_c^{(0)\tau n}(\lambda n)|pS_{\perp}\ra
\nonumber\\
&&+2f^{abc}\la pS_{\perp}|A^{\delta}_b(0)gA_c^{n}(0)F_a^{(0)\tau n}(\lambda n)|pS_{\perp}\ra
\Bigr]
S_{\mu\nu}(xp),
\label{part2}
\eeq
\beq
&&{1\over 2}\int{d^4k_1\over (2\pi)^4}\int{d^4k_2\over (2\pi)^4}
\int{d^4\xi}\int{d^4\eta}\,e^{ik_1\cdot \xi}e^{i\eta\cdot(k_2-k_1)}
\la pS_{\perp}|A_b^{\sigma}(0)gA_c^{\phi}(\eta)A_a^{\rho}(\xi)|pS_{\perp}\ra
\nonumber\\
&&\times \Bigl[
{if^{abc}\over k_2\cdot n-k_1\cdot n-i\epsilon}n_{\phi}{1\over k_1\cdot n-i\epsilon}
{1\over k_2\cdot n+i\epsilon}
\omega^{\mu}_{\ \tau}\omega^{\nu}_{\ \delta}(k_1^{\tau}n_{\rho}-(k_1\cdot n)g^{\tau}_{\ \rho})
(k_2^{\delta}n_{\sigma}-(k_2\cdot n)g^{\delta}_{\ \sigma})S_{\mu\nu}((k_2\cdot n)p)
\Bigr]
\nonumber\\
&=&{1\over 2}\int {dx\over x^2}\int{d\lambda\over 2\pi}
\,e^{i\lambda x}
\omega^{\mu}_{\ \tau}\omega^{\nu}_{\ \delta}
\Bigl[-if^{abc}\int^{\infty}_{\lambda}d\mu\la pS_{\perp}|F^{(0)\delta n}_b(0)igA_a^{n}(\mu n)
F_c^{(0)\tau n}(\lambda n)
|pS_{\perp}\ra
\nonumber\\
&&+if^{abc}\int^{\infty}_{\lambda}d\mu
\la pS_{\perp}|F^{(0)\delta n}_b(0)igA_c^{n}(\lambda n)F_a^{(0)\tau n}(\mu n)|pS_{\perp}\ra
\Bigr]
S_{\mu\nu}(xp),
\label{part3}
\eeq
\beq
&&{1\over 2}\int{d^4k_1\over (2\pi)^4}\int{d^4k_2\over (2\pi)^4}
\int{d^4\xi}\int{d^4\eta}\,e^{ik_1\cdot \xi}e^{i\eta\cdot(k_2-k_1)}
\la pS_{\perp}|A_b^{\sigma}(0)gA_c^{\phi}(\eta)A_a^{\rho}(\xi)|pS_{\perp}\ra
\nonumber\\
&&\times \Bigl[-{if^{abc}\over k_2\cdot n-k_1\cdot n-i\epsilon}n_{\phi}{1\over k_1\cdot n-i\epsilon}
{1\over k_2\cdot n+i\epsilon}
\omega^{\mu}_{\ \tau}\omega^{\nu}_{\ \delta}(k_1^{\tau}n_{\rho}-(k_1\cdot n)g^{\tau}_{\ \rho})
(k_2^{\delta}n_{\sigma}-(k_2\cdot n)g^{\delta}_{\ \sigma})S_{\mu\nu}((k_1\cdot n)p)
\Bigr]
\nonumber\\
&=&{1\over 2}\int {dx\over x^2}\int{d\lambda\over 2\pi}
\,e^{i\lambda x}
\omega^{\mu}_{\ \tau}\omega^{\nu}_{\ \delta}
\Bigl[if^{abc}\int^{\infty}_{0}d\mu\la pS_{\perp}|F^{(0)\delta n}_b(\mu n)igA_a^{n}(0)
F_c^{(0)\tau n}(\lambda n)
|pS_{\perp}\ra
\nonumber\\
&&-if^{abc}\int^{\infty}_{0}d\mu
\la pS_{\perp}|F^{(0)\delta n}_b(0)igA_c^{n}(\mu n)F_a^{(0)\tau n}(\lambda n)|pS_{\perp}\ra
\Bigr]
S_{\mu\nu}(xp).
\label{part4}
\eeq
Combining (\ref{part1})-(\ref{part4}), we obtain
\beq
&&\omega^{\mu}_{\ \tau}\omega^{\nu}_{\ \delta}\int {dx\over x^2}\int{d\lambda\over 2\pi}
\,e^{i\lambda x}
\Bigl[\la pS_{\perp}|F^{(0)\delta n}_b(0)\Bigl((-if^{abc})ig\int^{0}_{\lambda}d\mu\, A_a^{n}(\mu n)
\Bigr)F_c^{(0)\tau n}(\lambda n)
|pS_{\perp}\ra
\nonumber\\
&&+\la pS_{\perp}|F^{(0)\delta n}_b(0)gf^{bca}A_c^{\tau}(\lambda n)A_a^{n}(\lambda n)|pS_{\perp}\ra
+\la pS_{\perp}|gf^{abc}A^{\delta}_b(0)A_c^{n}(0)F_a^{(0)\tau n}(\lambda n)|pS_{\perp}\ra
\Bigr]
S_{\mu\nu}(xp).
\eeq
We find that they are $g^1$-terms of the gauge invariant form $\la pS_{\perp}|F^{\delta n}_b(0)
[0,\lambda n]_{ba}F^{\tau n}_a(\lambda n)|pS_{\perp}\ra$. The calculation for the first derivative term 
in the collinear expansion (\ref{collinear_twist-2}) can be simply carried by performing
additional partial integration with respect to $k_{1,2}$ in (\ref{part1})-(\ref{part4}). Thus we can show
\beq
&&i\omega^{\mu}_{\ \tau}\omega^{\nu}_{\ \delta}\omega^{\lambda}_{\ \kappa}
\int {dx\over x^2}\int{d\lambda\over 2\pi}
\,e^{i\lambda x}
\Bigl[\la pS_{\perp}|F^{(0)\delta n}_b(0)\Bigl((-if^{abc})ig\int^{0}_{\lambda}d\mu\, A_a^{n}(\mu n)
\Bigr)\partial^{\kappa}F_c^{(0)\tau n}(\lambda n)
|pS_{\perp}\ra
\nonumber\\
&&+\la pS_{\perp}|F^{(0)\delta n}_b(0)\Bigl((-if^{abc})ig\int^{\infty}_{\lambda}d\mu\, \partial^{\kappa}A_a^{n}(\mu n)
\Bigr)F_c^{(0)\tau n}(\lambda n)
|pS_{\perp}\ra
\nonumber\\
&&+\la pS_{\perp}|F^{(0)\delta n}_b(0)\partial^{\kappa}
\Bigl(gf^{bca}A_c^{\tau}(\lambda n)A_a^{n}(\lambda n)\Bigr)|pS_{\perp}\ra
+\la pS_{\perp}|gf^{abc}A^{\delta}_b(0)A_c^{n}(0)\partial^{\kappa}F_a^{(0)\tau n}(\lambda n)|pS_{\perp}\ra
\Bigr]
{\partial\over \partial k^{\lambda}}S_{\mu\nu}(k)\Bigr|_{k=xp},\hspace{5mm}
\eeq
which are $g^1$-terms of the matrix element of the kinematical functions (\ref{kinematical}).


\section*{Acknowledgments}

This research is supported by National Natural Science Foundation in China 
under grant No. 11950410495, Guangdong Natural Science Foundation under
No. 2020A1515010794 and research startup funding at South China
Normal University.

\end{document}